\documentclass{emulateapj}

\def\gtsim{\raise 2pt \hbox {$>$} \kern-1.1em \lower 4pt \hbox {$\sim$}}
\def\ltsim{\raise 2pt \hbox {$<$} \kern-1.1em \lower 4pt \hbox {$\sim$}} 
\defcitealias{gir04b}{Paper I}

\shorttitle{Sample of low-redshift BL Lac objects. II.}
\shortauthors{Giroletti et al.}

\begin{document}

\title{A sample of low-redshift BL Lacertae objects. II.  EVN and MERLIN data
and multi-wavelength analysis}

\author{M. Giroletti\altaffilmark{1,2},
        G. Giovannini\altaffilmark{1,2}, 
	G. B. Taylor\altaffilmark{3,4}, 
	R. Falomo\altaffilmark{5}       }

\altaffiltext{1}{INAF Istituto di Radioastronomia, via Gobetti 101, 40129,
Bologna, Italy; [giroletti,ggiovann]@ira.inaf.it}
\altaffiltext{2}{Dipartimento di Astronomia, Universit\`a di Bologna,
via Ranzani 1, 40127 Bologna, Italy}
\altaffiltext{3}{University of New Mexico, Dept. of Physics and
Astronomy, Albuquerque, NM 87131, USA; gbtaylor@unm.edu}
\altaffiltext{4}{National Radio Astronomy Observatory, P.O. Box O,
Socorro, NM 87801, USA}
\altaffiltext{5}{INAF Osservatorio Astronomico di Padova, vicolo Osservatorio
5, 35122, Padova, Italy; falomo@pd.astro.it}




\begin{abstract}

We present new radio observations of 9 members of a sample of 29 nearby ($z <
0.2$) BL Lac objects. The new data have been obtained with the European VLBI
Network and/or the MERLIN at 1.6 and 5 GHz and complement previous
observations. For one object, the TeV source Mrk~421, we also present deep
multi-epoch VLBA and Global VLBI data, which reveal a resolved diffuse jet,
with clear signatures of limb brightening. We use the new and old data to
estimate physical parameters of the jets of the sample from which the subset
with new radio data is drawn. We derive Doppler factors in the parsec scale
radio jet in the range $2 \la \delta \la 9$. Using HST data, we separate the
contribution of the host galaxy from that of the active core. From the measured
and de-beamed observables, we find a weak correlation between radio power and
black hole mass, and a tight correlation between radio and optical core
luminosities. We interpret this result in terms of a common synchrotron origin,
with little contribution from a radiatively efficient accretion disk. The BL
Lacs in our sample have de-beamed properties similar to low power radio
galaxies, including the fundamental plane of black hole activity.

\end{abstract}

\section{INTRODUCTION}

In the Unification schemes of radio loud AGN, the reason for the difference
between high power (FR II radio galaxies and quasars) and low power (FR Is and
BL Lacertae objects) sources is still a puzzling subject. In particular, it is
not yet clear if it is related to evolutionary properties, to differences in
the physical state of the central black hole \citep[BH, see e.g.][]{bau95}
and/or to a different interaction between the relativistic jet and the
interstellar medium \citep{dey93}.

In this context, the study of BL Lac sources is important for many reasons.
First, BL Lacs are not well studied in the radio band and their physical
properties in this band are not yet well known. For instance, detailed VLBI
studies on BL Lacs are available only for the bright objects of the 1 Jy sample
\citep{cas02} and a few other outstanding sources
\citep{gir04a,pin04,rec03}. In spite of this lack of high resolution data, BL
Lacs are ideal objects to be observed with VLBI, since they are beamed objects
and represent a good laboratory to look for evidence of the presence of
relativistic jets in low power radio sources \citep{den00}.

In order to investigate this class of AGN, we selected a sample of nearby BL
Lacs from the Hubble Space Telescope (HST) snapshot image survey
\citep{urr00,sca00}. The sample is composed of 30 objects with redshift $z<0.2$
\citep[~hereinafter Paper I]{gir04b}. At these low redshifts, the HST separates
the contribution of the central core and of the host galaxy; in the radio, we
can probe with VLBI the parsec scale relativistic jet. Moreover, we can study
low power objects that are weak in the radio, with a spectral energy
distribution (SED) peaking at high $\nu_{\rm peak}$ (HBLs); in the inverse
Compton part of their SED, some members of the sample reach even the TeV
domain. By contrast, their radio power is extremely low, which makes them
comparable to radio quiet objects, when we consider their de-beamed luminosity.

In \citetalias{gir04b}, we presented VLA and VLBA observations for 15
objects. We have discussed some of the basic radio properties of the sample and
we have made a first estimate of the parsec scale jet properties. In the
present paper we present new observations of nine objects with the European
VLBI Network (EVN) and the MERLIN, including unpublished images of
Mrk~421. Moreover, we take advantage of the data from the HST snapshot survey
and from X-ray satellites, in order to discuss the properties of the sample
from a broadband perspective.

The paper is laid out as follows: \S\ref{sample} contains a summary of the
results of \citetalias{gir04b} and introduces the new data-sets; in
\S\ref{results}, we present the results of the new observations and details
about individual sources; in \S\ref{discussion}, we discuss the broadband
properties of the sample; finally, \S\ref{conclusions} summarizes the main
conclusions of the present work. Throughout the paper, we make use of H$_0$ =
70 km sec$^{-1}$ Mpc$^{-1}$, $\Omega_M$ = 0.3 and $\Omega_\Lambda$ =
0.7. Spectral indexes are defined such that $S(\nu) \propto \nu^{-\alpha}$.

\section{SAMPLE PROPERTIES AND NEW OBSERVATIONS}
\label{sample}

\subsection{The sample: a summary}

Based on the HST snapshot survey of BL Lac objects \citep{urr00,sca00}, we have
selected a sample of 29 BL Lac objects with low redshift ($z <0.2$). We point
out that the sample considered in \citetalias{gir04b} was composed of 30
sources. However, thanks to our analysis, we have been able to discard one
object from the sample, that is not a genuine BL Lac: we reject 0145+138 from
the sample, on the basis of its low core dominance ($S_\mathrm{c}/S_\mathrm{t}
= 0.08$) and of the lack of an unresolved nuclear component in the optical HST
image. More observations and an optical spectrum would be desirable in order to
understand its true nature. We also note that some objects (e.g. 0521$-$365,
2201+044) have optical spectra showing emission and absorption lines and are
similar to Seyfert 1 spectra but with lines of lower luminosity. In total, we
have 7 sources with low frequency synchrotron peak (LBL), and 22 HBLs. The
total radio powers span the range 10$^{23.9}$ to 10$^{26.1}$ W Hz$^{-1}$ at 1.4
GHz.

Table \ref{table1} shows in detail the list of objects in the sample and their
basic properties: IAU name and other name in cols.~[1] and [2], redshift in
col.~[3], monochromatic power at 1.4 GHz from the NVSS in col.~[4], optical
absolute magnitude in $R$ band from the HST snapshot survey \citep{urr00}
in col.~[5], and X-ray luminosity in the 2-10 keV band in col.~[6],
obtained from the BeppoSAX archive cataloged by \citet{don05}, or from other
references as listed in col.~[7].  Note that the radio and X-ray emission in BL
Lacs is variable; when more than one measurement was available in the BeppoSAX
archive, we used the mean value.  Columns [8] and [9] report the viewing angle
$\theta$ and Doppler factor $\delta$ estimated on the basis of the arguments
discussed in \citetalias{gir04b}\footnote{In \citetalias{gir04b}, we only
listed values assuming $\Gamma = 5$. The data shown in cols.~[8] and [9] are
derived in the case that $\Gamma \sim 1/\theta$ and they can then be different
from those found in our earlier work.}.  Finally, we report the object type in
col.~[10] and the original sample in col.~[11].

In \citetalias{gir04b}, we presented radio images from pc to kpc scales for the
whole sample, considering new and published VLA and VLBA observations. All
objects are core dominated on kiloparsec scales; extended structure is however
usually present, either symmetric or one-sided, in the form of jets, halos,
and/or secondary compact components. On the parsec scales, we found weak cores
and a few short, one-sided jets. The VLBA images account for most of the
arcsecond core flux density, implying that we have not missed considerable
subarcsecond structure. Parsec- and kiloparsec-scale jets are oriented at the
same P.A. in a large fraction of objects, in contrast with the large
bending angles found in several BL Lacs of the 1 Jy sample
\citep[$\Delta\mathrm{P.A.}>90^\circ$, ][]{cas02}. If large bends are ascribed
to geometric amplification of small oscillations in jets closely aligned to the
line of sight, the similarity in the P.A.~of the parsec and kiloparsec scale
jets found in our sample implies that their viewing angles are larger than
those of the LBLs in the 1 Jy sample \citep[see also][]{rec03}. From the
jet/counterjet ratio, the synchrotron self-Compton (SSC) model, and primarily
from the core dominance, we estimate the range of possible parameters of the
jets. Under the statistical assumption of $\Gamma \sim 1/\sin\theta$, we
derived an average viewing angle of $\langle \theta \rangle = 18^\circ \pm
5^\circ$ and Lorentz factors $\Gamma$ of the jets in the range $2 \la \Gamma
\la 9$.

The properties of the host galaxies have been presented by \citet{fal00}, based
on the original HST data for the sample. The hosts are virtually all massive
elliptical galaxies (average luminosity $M_\mathrm{R} = -22.9 \pm 0.6$ and
effective radius R$_e$ = 10 kpc) typically located in poor groups. In spite of
the presence of the active nucleus, the host galaxy appears in most cases to be
a completely ``normal'' elliptical galaxy. Moreover, the data have shown that
no significant difference is present between the hosts of HBLs and those of
LBLs.

\subsection{Observations and data reduction}
\label{observations}

In this paper we present new radio observations of nine BL Lacs. We list the
various data set at our disposal in Table \ref{tab2}. For 8 sources (i.e. all
except 1101+384), we have European VLBI Network (EVN) observations at 1.6 GHz
in single polarization taken on 2002 June 7th. The stations involved were
Effelsberg (100 m), Jodrell Bank (MkII, $25$ m), Medicina (32 m), Noto (32 m),
Onsala (25 m), Sheshan (25 m), Torun (32 m), Westerbork ($14 \times 25$ m), and
Urumqi (25 m). Each source was observed for $\sim 55$ minutes, divided into
four scans at different hour angles to obtain a good $(u,v)$ coverage. The data
were correlated at the Joint Institute for VLBI in Europe (JIVE) in
Dwingeloo. We imported the data in the National Radio Astronomy Observatory
(NRAO) Astronomical Image Processing System (AIPS) and performed the a priori
calibration. Finally, we edited the data and produced the images within {\sc
Difmap}; the final restoring beams, 3$\sigma$ noise levels, and image peaks are
reported in Table \ref{imagedata}.

Two of the sources (1215+303 and 1728+502) were also observed at 5 GHz with the
EVN and the MERLIN simultaneously on 2002 May 26th. The EVN array involved in
the experiment includes the same telescopes of the EVN only observation, plus
the 32~m antenna in Cambridge; this antenna provides a baseline to Jodrell
Bank, common to the MERLIN array as well. Each source was observed for 6 hours,
in 42 scans, switching between the two targets. The MERLIN data have been
processed at Jodrell Bank Observatory with the local {\tt dprograms}; {\sc
Difmap} has been used to produce the images. The EVN data were correlated in
Dwingeloo and reduced in the standard manner within AIPS (for initial
calibration) and {\sc Difmap} (for the imaging part). Finally, the combination
of the two dataset has been done within AIPS, using the task {\tt DBCON}. For
the combined dataset, we have modified the weights of the original data, in
order to optimize the compromise between resolution and sensitivity, by means
of the adverb {\tt REWEIGHT} in {\tt DBCON}; in particular, the EVN data have
been rescaled by a factor of 20.

Moreover, we present here unpublished data for 1101+384 (Markarian 421): (i)
EVN+MERLIN observations taken on 1997 Feb 17 at 5 GHz: the EVN telescopes
involved in the observation were Effelsberg, Sheshan, Cambridge, Medicina,
Torun, and the VLBA dish in Saint Croix; the MERLIN array consisted of Defford,
Cambridge, Knockin, Darnall, MkII at Jodrell Bank, and Tabley; we have about
six hours of EVN observations and three hours with the MERLIN, after flagging
for down time, bad weather, receiver problems; we followed the procedures
outlined above for the data reduction of the EVN and MERLIN data set and for
their combination; (ii) observations with the VLBA plus a single VLA dish
(VLBA+Y1) for 4 hours at 5 GHz on 1995 July 22; (iii) observations at 1.6 GHz
on 1996 February 10 with a global array (6 VLBA telescopes for 6 hours,
including 3 hours at Jodrell Bank, Medicina, Noto, and Onsala). A preliminary
map based on this observation was presented by \citet{gio99}. See Table
\ref{merlindata} for the restoring beam, peak and noise level of all our new
images of Mrk~421.

\section{RESULTS}
\label{results}

\subsection{General Properties}

\begin{figure}
\plotone{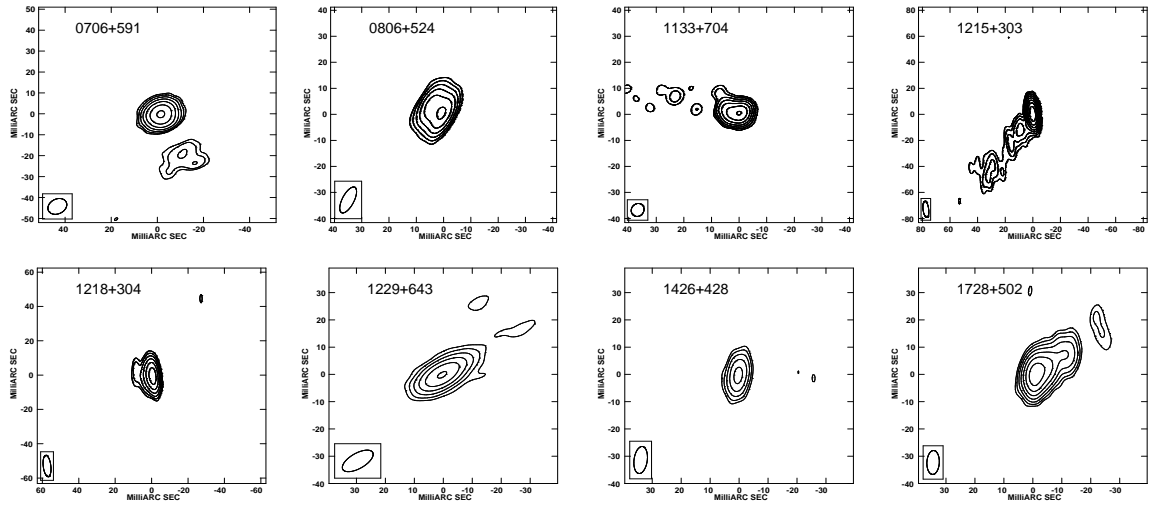}
\figcaption{EVN images taken at 1.6 GHz on 2002 June 7. Contours are
drawn at (1, 2, 4, 8, 16, ...) times the noise level. Noise levels and
image peaks are given in Tab. \ref{imagedata}.
\label{fig1} }
\end{figure}

We show in Fig.~\ref{fig1} the milliarcsecond scale images of the eight BL Lacs
observed at 1.6 GHz with the EVN; we report their relevant parameters in
Table~\ref{imagedata}. Typically, the sources are dominated by a compact core
making up the largest fraction of the total flux density. Jets are detected in
seven out of eight sources, always single sided. This rate of jet detection
(83\%) is higher than in the 5 GHz VLBA observations of other objects of the
sample \citepalias[8/15, or 53\%, see][]{gir04b}. This may stem from both the
longer exposure and brighter targets of the present observations.

We also present several deep radio images at 5 GHz for 1215+303
(Fig.~\ref{fig1215}) and 1728+502 (Fig.~\ref{fig1728}), and at 1.6 and 5 GHz
for 1101+38 (Mrk~421, Figs.~\ref{fig1101a},~\ref{fig1101b}). These images span
scales from a few parsecs to some kiloparsecs, and include our new EVN+MERLIN
combination. These sources possess bright cores and longer ($> 100$ mas)
one-sided jets. The jets are remarkably well aligned on scales from a few to
some hundreds of parsecs (projected). Evidence of bending is present only in
the inner parsecs of the jet of Mrk 421. We provide more details on these
sources in \S\ref{notes},\ref{notes1101}, including a description of their
kiloparsec scale properties.

In Table~\ref{modelfits}, we report the results of our modelfits to the
visibility data for the EVN 1.6 GHz observations. In Col. (1) we give the
source name, in Col. (2) the various components studied for each source, and in
Cols. (3-6) the details of the model for that component: flux density, radius
$r$ (i.e. distance from an arbitrary origin), position angle $\theta$, and FWHM
$b$. These models have been obtained fitting circular Gaussian components to
the data using modelfit in Difmap. We adopt circular Gaussians because the
signal to noise ratio is relatively low in most jets and we need to keep the
number of parameters as low as possible. The models for the most compact and
powerful components are typically better constrained than for weak and extended
ones, although the various model parameters are not independent of each
other. Typically, components smaller than 10 mas have uncertainties on $r$ and
$b$ of $\sim0.5$ mas or less. For larger components, uncertainties are
estimated on the basis of the HPBW of the image and the component SNR. For the
flux densities, we estimate the uncertainty from the rms noise and the absolute
flux calibration uncertainty ($\sim10\%$) added in quadrature.  We then
summarize in Table~\ref{tab407} the main results about cores and jets
properties, based on the new 1.6 and 5 GHz EVN data, as well as on the 5 GHz
VLBA data presented in \citetalias{gir04b}. We give the core flux density at
1.6 GHz in Col. (2), and the one at 5 GHz in Col. (3), either using present EVN
data or the VLBA data in \citetalias{gir04b}. The corresponding
non-simultaneous spectral index is given in Col. (4). Columns (5) and (6)
contain the jet brightness $B_{\rm j}$ and the relative lower limit on the
jet/counter-jet ratio $R_{\rm min}$.

Core flux densities have been derived from the fits of Gaussian components to
the data. The core flux density is expected to be slightly higher than the peak
brightnesses of the images reported in Table~\ref{imagedata} because of the
discrete map gridding. In a few cases (1133+704, and 1215+303), the
significantly larger core flux density indicates that there is some
contamination from the jet. At full resolution, the different observing
frequency brings about an excess of flux density in the 1.6 GHz data, mostly as
a result of the different size of the synthesized beams. For a spectral index
measurement, we therefore consider our 5 GHz images (either VLBA or EVN),
limiting the $(u,v)$-range to $0-40$ M$\lambda$; this yields a resolution
comparable to the 1.6 GHz images. Results are presented in Table \ref{tab407},
with $\alpha_{1.6}^{5}$ in the range between $-0.1$ and 0.5. Note that the
observations at 1.6 and 5 GHz are not simultaneous, so variability can account
for part of the differences.

All the sources display single-sided jets in both 1.6 and 5 GHz data, except
for 1426+428, which has a tentative jet detection only at 5 GHz
\citep{kol96}. In most cases, we need only one component (in addition to the
core) to model-fit the jet. We also measured flux densities and brightnesses
directly on the images.  Since no object in the sample has a detectable
counter-jet, the brightness ratios presented in Table~\ref{tab407} are only
lower limits based on the noise of the images. More details are given along
with the source description in the next section.

For 1215+303, 1728+502, and Mrk~421 we have also obtained spectral index maps
(simultaneous in the case of 1215+303 and 1728+502). Cores are typically
flat/inverted, while the jets are steeper in general. However, the
signal-to-noise is too low to make any strong statement about possible
substructures within the jet and we do not show these images here.

\subsection{Notes on single sources}
\label{notes}

{\it 0706+591 --} The EVN jet points at $\sim -150^\circ$. A significant
fraction of flux density (about 14\%) is found in a component at $\sim 17$ mas
($\sim 40$ pc) from the core, with a peak brightness of 2.0 mJy/beam. The
comparison with VLA and VLBA images in \citetalias{gir04b} \citep[see
also][]{gir05} confirms the orientation of the parsec scale jet and its
misalignment to the kiloparsec scale extended structure. The core sits at the
northwest edge of a diffuse extended halo, with major axis oriented in
P.A. $\sim 120^{\circ}$.

{\it 0806+524 --} The EVN observation at 1.6 GHz recovers a total of 181
mJy. This source is then one of the most core dominated in the sample
($S_\mathrm{NVSS} = 183$ mJy). A well defined jet component is found at 4.8 mas
from the core in P.A.  $42^\circ$, i.e. only slightly misaligned
($\Delta$P.A.$\sim 35^\circ$) with respect to the VLBA jet at 5 GHz
\citepalias{gir04b}.

{\it 1101+384 (Mrk 421)--} See \S\ref{notes1101}

{\it 1133+704 --} Our data reveal a short inner jet (P.A.$_\mathrm{j} \sim
70^\circ$) and some more distant low brightness emission. Both features are
well aligned with the major axis of the kiloparsec scale halo shown in
\citetalias{gir04b}, as well as the parsec scale jet found by \citet{kol96}.

\begin{figure}
\plotone{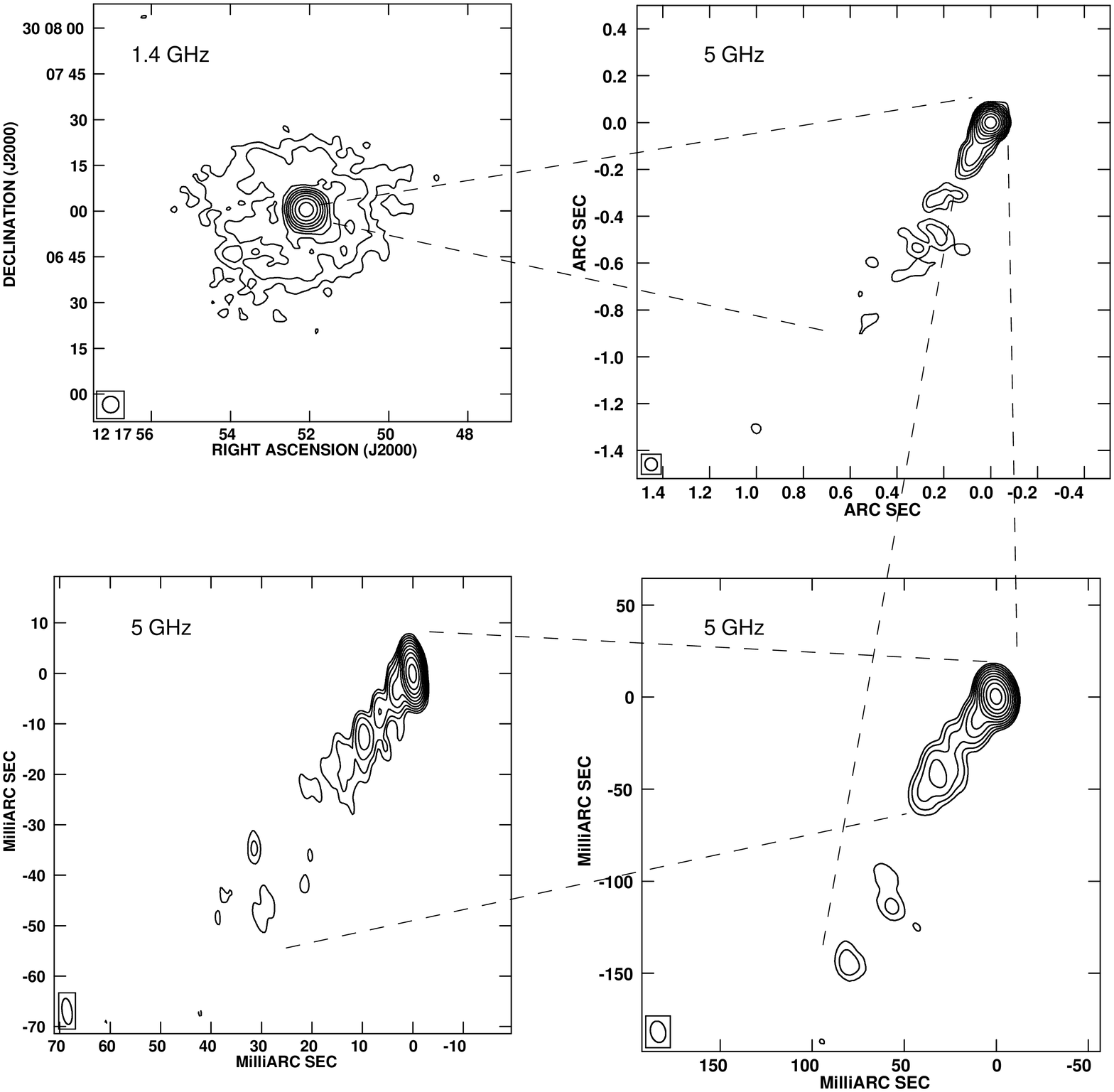}
\figcaption{Images of 1215+303. Clockwise from top left: FIRST, MERLIN, EVN+MERLIN, and EVN only images. Contours are traced at ($-1$, 1, 2, 4, ...) times the noise levels. Beam, noise and peak for the EVN and MERLIN images are given in Table \ref{merlindata}; the noise in the FIRST image is 0.45 mJy/beam.
\label{fig1215} }
\end{figure}

\begin{figure}
\plotone{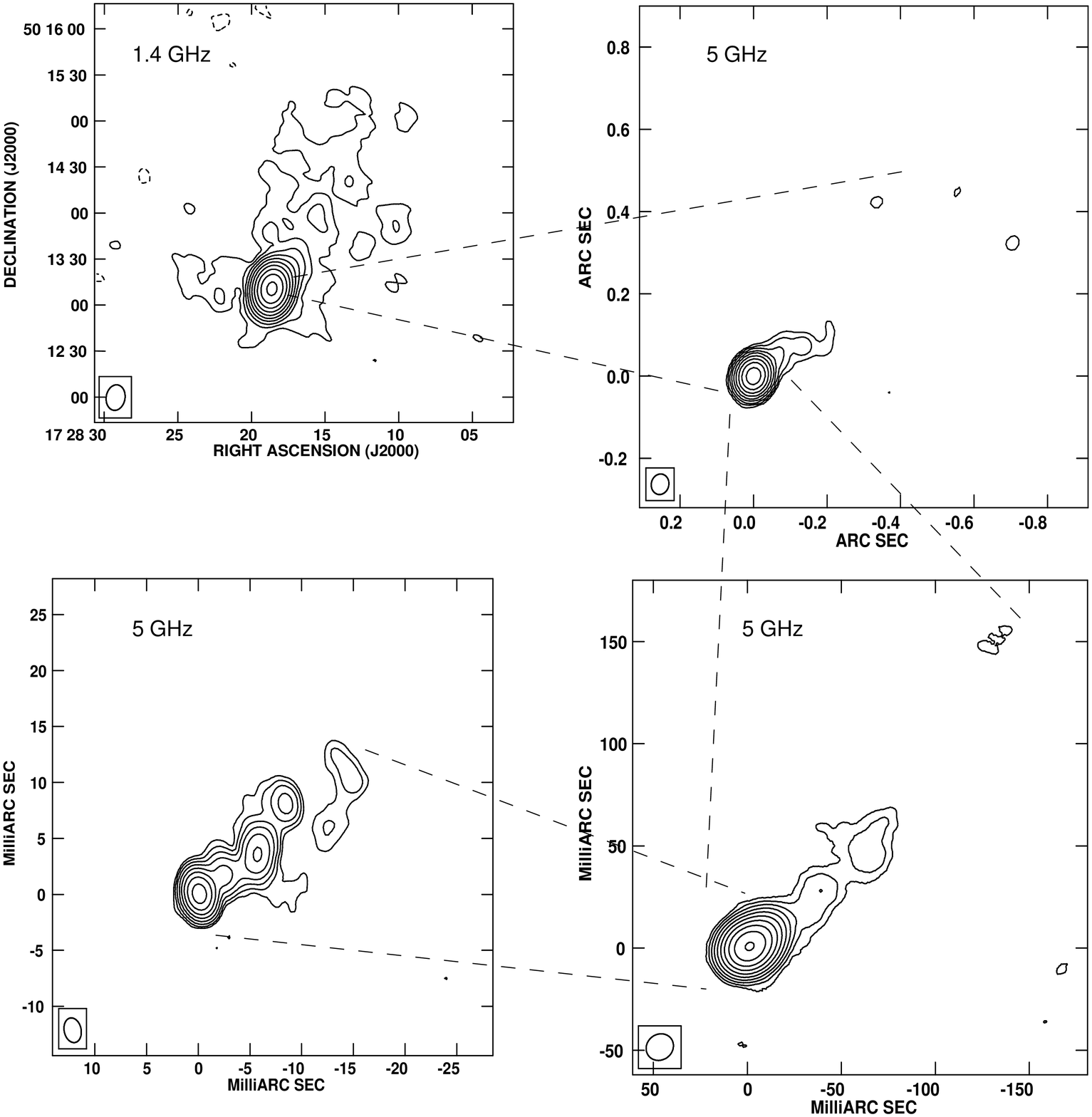}
\figcaption{Images of 1728+502. Clockwise from top left: VLA, MERLIN, EVN+MERLIN, and EVN only images. Contours are traced at ($-1$, 1, 2, 4, ...) times the noise levels. Beam, noise and peak for the EVN and MERLIN images are given in Table \ref{merlindata}; the noise in the VLA image is 0.3 mJy/beam (see \citetalias{gir04b}).
\label{fig1728} }
\end{figure}

\begin{figure}
\plotone{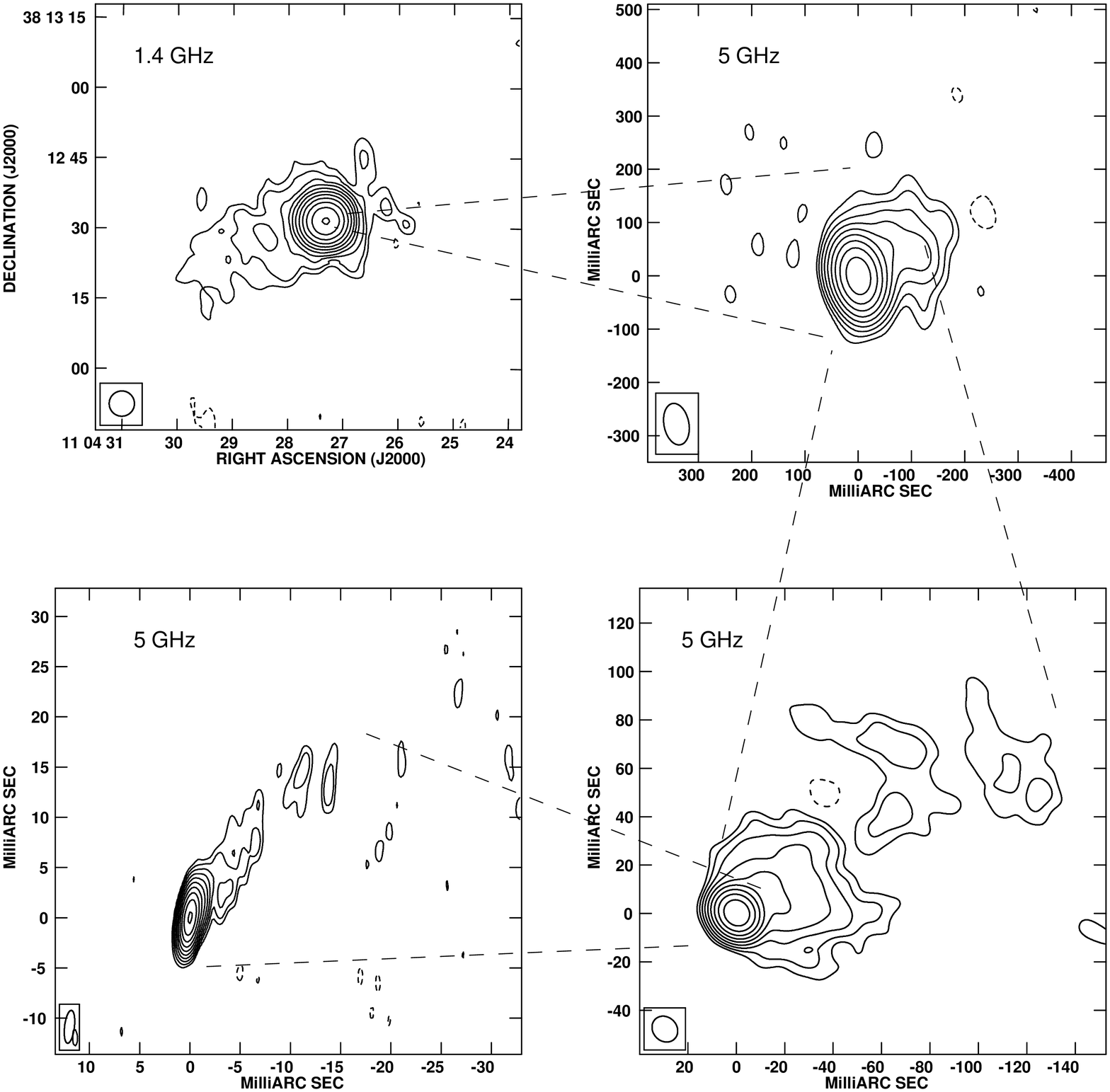}
\figcaption{Images of Mrk 421. Clockwise from top left: FIRST, MERLIN, EVN+MERLIN, and EVN only images. Contours are traced at ($-1$, 1, 2, 4, ...) times the noise levels. Beam, noise and peak for the EVN and MERLIN images are given in Table \ref{merlindata}; the noise in the FIRST image is 0.5 mJy/beam.
\label{fig1101a} }
\end{figure}

{\it 1215+303 --} The HBL 1215+303 is located at $z=0.130$. For this object, a
good map is available in the FIRST survey \citep{bec95}, as shown in
Fig. \ref{fig1215} (top left panel). A 377 mJy core dominates a symmetric halo
structure of about 50'' diameter. No preferred direction is visible on this
scale, while the EVN 1.6 GHz image shows a jet oriented in PA $140^\circ$. For
this reason, we consider also the EVN+MERLIN observations at 5 GHz, in order to
image the inner structure out to the largest possible distance from the inner
core. The other panels in Fig. \ref{fig1215} show the long, straight, one-sided
jet. The large jet/counter-jet ratio ($R\ga 150$) indicates that the effects of
the beaming are important in this source; indeed, one could expect a small
angle to the line of sight, if the kpc halo structure were interpreted as a
lobe seen end-on. 

{\it 1218+304 --} The total VLBI correlated flux at 1.6 GHz is 64 mJy and it
dominates the total flux density of the source ($S_\mathrm{NVSS} = 71.3$
mJy). The core spectral index is $\alpha = 0.1$; a weak jet in P.A. $84^\circ$
is present at both 1.6 and 5 GHz \citepalias[see][]{gir04b}.

{\it 1229+643 --} The 5 GHz VLBA image \citepalias{gir04b} clearly shows a
core-jet structure in P.A.$ = - 43^{\circ}$. In the present 1.6 GHz image, most
of this emission is blended in the core component. Little flux density is
detected in the same position angle at a larger distance. The total flux
density recovered by the EVN is 61 mJy and exceeds that of the kiloparsec scale
core \citepalias[55 mJy,][]{gir04b}, calling for significant variability in
this source.

{\it 1426+428 --} This BL Lac is a TeV source \citep{aha02,hor02}. It is
unresolved in our EVN image, with a total flux density of 32 mJy. This makes up
about 50\% of the total flux density at 1.4 GHz in the NVSS (61 mJy).

{\it 1728+502 --} This source displays a remarkable alignment from the parsec
scale EVN image out to the NVSS map, going by our new MERLIN and earlier VLA
\citep{gir04b} observations (see Fig.  \ref{fig1728}). The parsec scale jet in
PA $-55^\circ$ is still collimated on the MERLIN scale, with a little tilt
east. Then, in the C array VLA image, the emission is spread over more than
100", for a total of $\sim 16$ mJy of extended emission. This emission is
resolved in the A array image, which is dominated by the 200 mJy core.

\subsection{Markarian 421}
\label{notes1101}

\begin{figure}
\plottwo{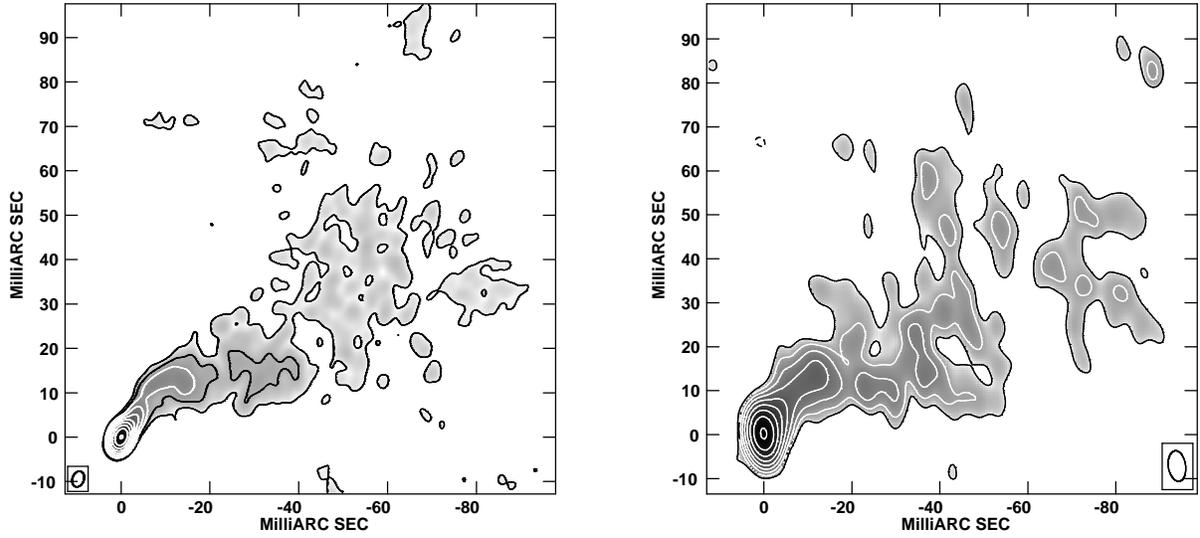}{f5b.eps}
\figcaption{Images of Mrk 421. Left: VLBA+Y1 at 5 GHz (epoch 1995.7); right: Global VLBI at 1.6 GHz (epoch 1996.2). Contours are traced at ($-1$, 1, 2, 4, ...) times the noise levels. Beam, noise and peak are given in Table \ref{merlindata}. 
\label{fig1101b} }
\end{figure}

\begin{figure}
\epsscale{0.5}
\plotone{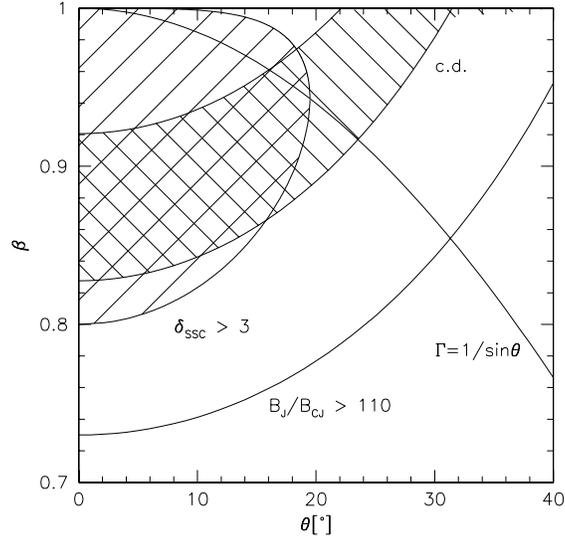}
\figcaption{Plane $(\beta, \theta)$ for the jet of Mrk~421, with constraints derived from jet/counterjet ratio, core dominance (c.d.), and synchrotron self-Compton model. The cross-hatched area represents the  $(\beta,\theta)$ pairs that satisfy the observational data. The relation $\Gamma=1/\sin\theta$ is also shown; its section within the hatched area corresponds to the most likely combinations of jet velocity and orientation.  \label{bth} }
\epsscale{1.0}
\end{figure}

Markarian 421 ($z=0.031$, 0.6\arcsec/kpc) is a well known BL Lac, widely
studied at all frequencies and detected at TeV energies \citep{pun92}. The NVSS
image shows a 30\arcsec\, core-dominated source, with emission on either
side. As we increase the resolving power of our images and get closer to the
core region, more and more extended emission becomes resolved out and finer
details are discovered.  In Fig.~\ref{fig1101a}, we see the dominant structure
shift from the eastern side (FIRST image, top left panel) to the west (MERLIN
and EVN images, clockwise from top right). The parsec scale data (see also
Fig.~\ref{fig1101b}) convincingly show that the approaching side is the western
one. On the basis of the parsec scale observational data, we estimate that the
viewing angle is $\theta \la 20^\circ$. In Fig.~\ref{bth}, we combine the
constraints posed by the jet-counterjet ratio ($R > 110$, at 13 mas from the
core), core dominance, and synchrotron self-Compton excess,
$\delta_\mathrm{SSC} \ga 3$, based on a VLBI core of $\sim 300$ mJy, and an
upper limit to the angular diameter of 0.1 mas. We estimate an
orientation for the radio jet of $\theta = 18^\circ \pm 2^\circ$, under the
assumption of $\Gamma \sim 1/\sin\theta$.  This corresponds to the maximum
angle allowed for a given Doppler factor, which is, in a statistical sense, the
most likely situation.  Such relatively large angle for a TeV radio source may
indicate that this assumption could be inaccurate for Mrk~421; however, it
could also be suggestive of a change in the inner jet orientation, similarly
to what is found in Mrk~501 \citep{gir04a}, as well as of the presence of a
structured jet \citep[see also][]{ghi05}.

The jet of Mrk~421 is bright and well-collimated in the inner $\sim 20$ mas,
and it then becomes fainter and wider, with an opening angle $\phi_o \sim
40^\circ$. Under the viewing angle of $18^\circ$, this corresponds to an
intrinsic jet opening angle $\phi_i \sim 12^\circ$. The jet shows also some
bending from P.A.~$-35^\circ$ close to the core to P.A.~$-65^\circ$ at larger
distance. Both the change in collimation and the bending -- though less
conspicuous -- remind us of the other TeV BL Lac Mrk 501, whose parsec scale
jet turns by $55^\circ$ and widens significantly at about 30 mas from the
core. \citet{gir04a} have studied the jet of Mrk~501, discussing also evidence
of limb brightening in the inner 30 mas.

\begin{figure}
\plottwo{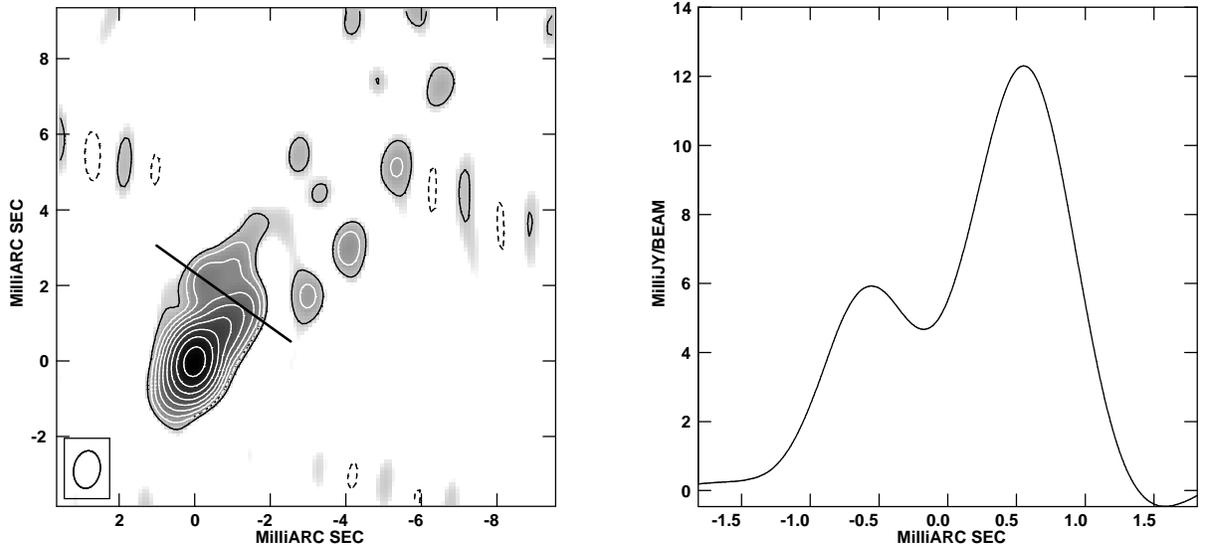}{f7b.eps}
\figcaption{Left: high resolution (beam $1.0\times 0.7$ mas in P.A.~$-9^\circ$) image of the inner jet of Mrk~421 at 5 GHz with EVN, contoured at $(-1, 1, 2, 4, ... ) \times 0.86$ mJy/beam; right: brightness profile along the slice shown in the left panel.
\label{lb1} }
\end{figure}

\begin{figure}
\epsscale{0.3}
\plotone{f8a.eps}
\plotone{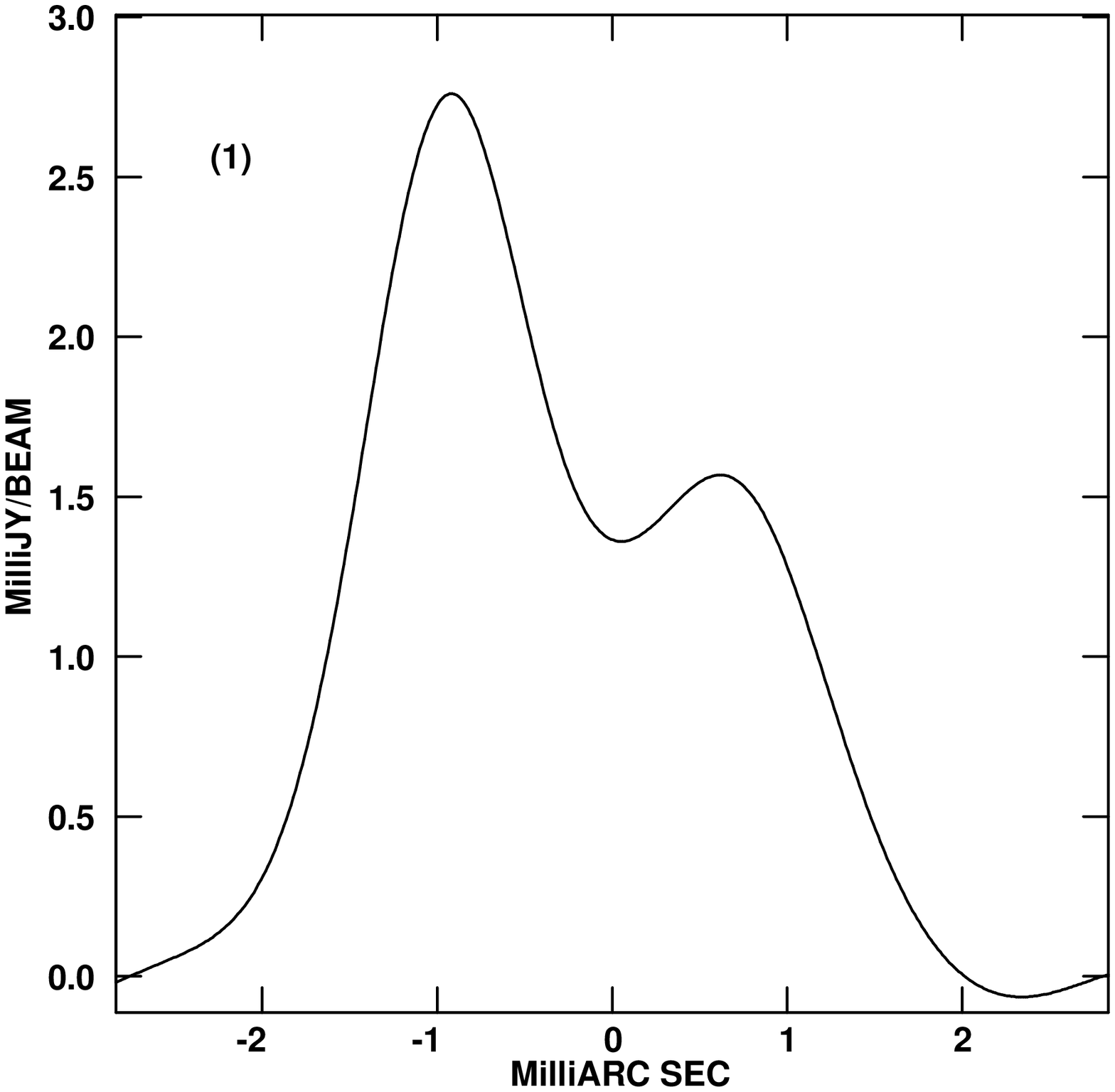}
\plotone{f8c.eps}
\epsscale{1.0}
\figcaption{Left: high resolution (beam $2\times 1$ mas in P.A. $-21^\circ$) image of the inner jet of Mrk~421 at 5 GHz with VLBA+Y1, contoured at $(-1, 1, 2, 4, ... ) \times 0.3$ mJy/beam; the brightness profiles along the slices are shown in the middle and right panels.\label{lb2} }
\end{figure}

Although less prominent, an enhanced brightness at the jet edge is visible also
in our images of Mrk~421. In Fig.~\ref{lb1} we show a high resolution image
obtained from the 1997 EVN data with uniform weights: the jet is clearly
transversally resolved at $\sim 2$ mas from the core \citep[see
also][]{pin05}. Thanks to the 1995 VLBA+Y1 5 GHz observation, we extend this
study and consider a more distant region, at $\sim 10-15$ mas, which is
illustrated in Fig.~\ref{lb2}. The grey-scale image and the brightness profiles
across the jet reveal a clear limb-brightening, which is most prominent on the
north-east side. After the bend, the jet is well resolved and it is clearly not
centrally peaked; a possible limb-brightening could also be present at about 30
mas from the core on the right edge of the jet, as visible in
Fig.~\ref{fig1101b} both at 1.6 and 5 GHz.

If we interpret the brightness morphology as a velocity structure in the jet,
we have an agreement with observational data if $\theta=18^\circ$ assuming a
fast spine structure ($\Gamma_\mathrm{spine}=10-15$, corresponding to
$\delta=1.9-1.3$) and a slower external sheath ($\Gamma_\mathrm{sheath}=2-5$,
corresponding to $\delta=2.9$). The larger $\delta$ in the slower sheath will
thus account for the observed brightness structure.

As is the case also in Mrk~501, the resolved and limb brightened structure of
the jet makes it difficult to describe the jet with simple Gaussian
components. We have nonetheless tried to model-fit the data with circular
Gaussians, and we find that the inner jet is well described by a core plus
three components (see Table~\ref{tab421}), in good agreement with results at
higher frequency of \citet{pin05}. In particular, we match our two innermost
components with their C4 and C5. However, given the larger number of epochs
considered by \citet{pin05}, our data do not improve over their study of the
motion of component, which we regard as conclusive. The small observed
velocities (peak velocity of $(0.1 \pm 0.02)c$) confirm also our results on the
similar source Mrk~501 \citep{gir04a}: low power radio sources with complex,
limb-brightened jets show low $\beta_\mathrm{pattern}$, which is therefore not
related the bulk velocity of the jet.

\section{Discussion}
\label{discussion}

Thanks to the new data presented here and in \citetalias{gir04b}, we estimate
the relevance of beaming in our sample of nearby BL Lacs. As discussed in
\citetalias{gir04b}, we base our estimates for the jet parameters on (1) the
jet-counterjet brightness ratio; (2) the core enhancement with respect to the
value expected from the low frequency, not boosted total radio power; (3) the
measured 1 keV flux densities with respect to the SSC predicted ones. In
particular, the core dominance selects a region of the space of parameters
$(\beta, \theta)$, which can be either fully consistent or slightly restricted
by the limits posed by (1) and (3). From this allowed region of the space of
parameters, we derive the Lorentz factor $\Gamma$ and the viewing angle
$\theta$, under the assumption that $\Gamma \sim 1/\sin\theta$. We note that
for any one source the relation may not hold, although in a statistical sense
this shall be the most likely solution.

With this caveat in mind, we derive the intrinsic properties of the
sources in our sample. In this way, we are able to discuss their physical
properties and address the issue of the parent population. For example, we have
shown in \citetalias{gir04b} that the de-beamed nuclear radio luminosity are
indistinguishable from those of a sample of FR~I radio galaxies, in addition to
what is already well known about the extended radio luminosity. We will now
adopt a broadband perspective and consider multi-wavelength data, which provide
a tool to investigate the mechanisms involved in the nuclear activity. For the
present sample, a wealth of information is available thanks to the optical
observations performed with the Hubble Space Telescope. In particular, for all
the objects it has been possible to separate the contribution of a central
compact core and of the host galaxy. Both quantities have significant impact on
our understanding of the BL Lac phenomenon. The central compact core is a clear
signature of the nuclear activity; the host galaxy luminosity is a tool to
estimate the mass of the central black hole \citep{mag98,kor01}.

\subsection{BH mass and radio power}
\label{radiooptical}

In the following, we adopt black hole mass estimates based on the host galaxies
bulge luminosity. As discussed by \citet{fct03}, this method is preferable to
those based on the correlation with the central stellar velocity dispersion
$\sigma$ \citep{fer00,geb00}. In fact, (1) direct measurements of $\sigma$ in
BL Lacs are difficult to obtain and (2) indirect estimates of $\sigma$ via the
fundamental plane (FP) are highly inaccurate due to the intrinsic scatter of FP
relation, as well as to the difficulties in determining the other observables
in the FP.

\begin{figure}
\epsscale{0.5}
\plotone{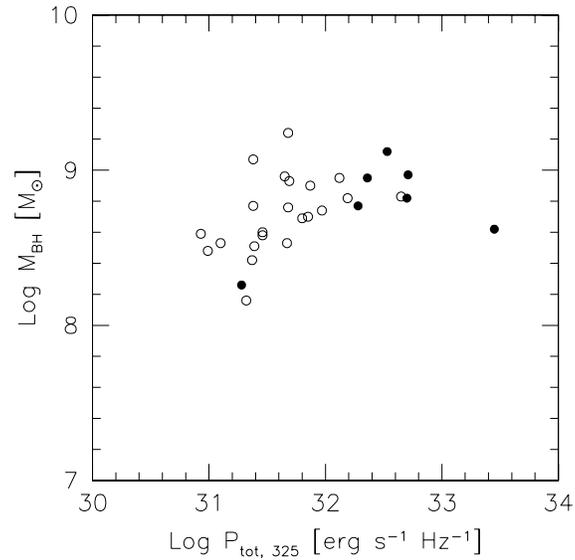}
\figcaption{Black hole mass vs. total radio power at 325 MHz. Filled and empty
  symbols represent HBL and LBL respectively.
\epsscale{1.0}
\label{fig:bh1} }
\end{figure}

\begin{figure}
\plottwo{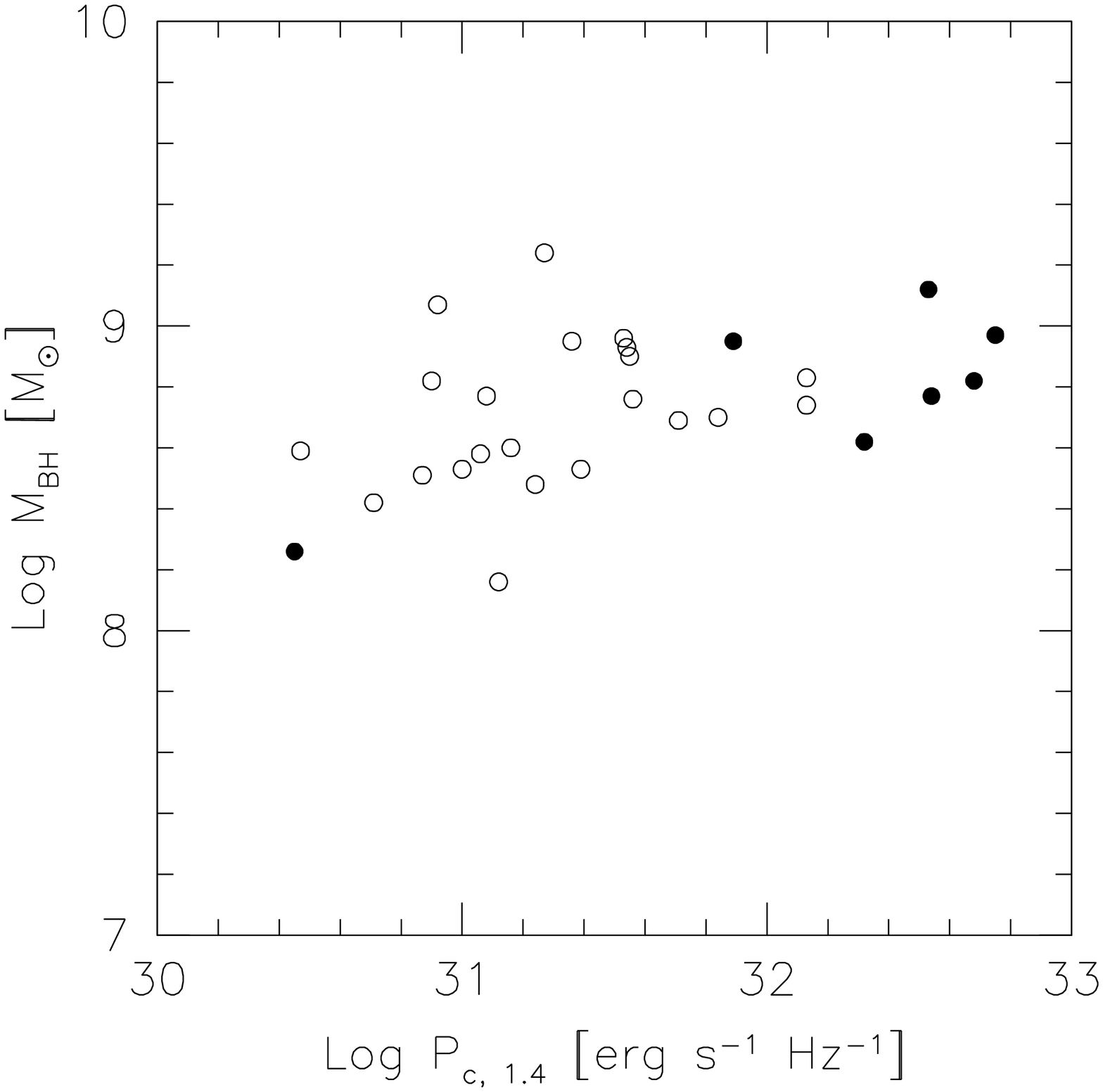}{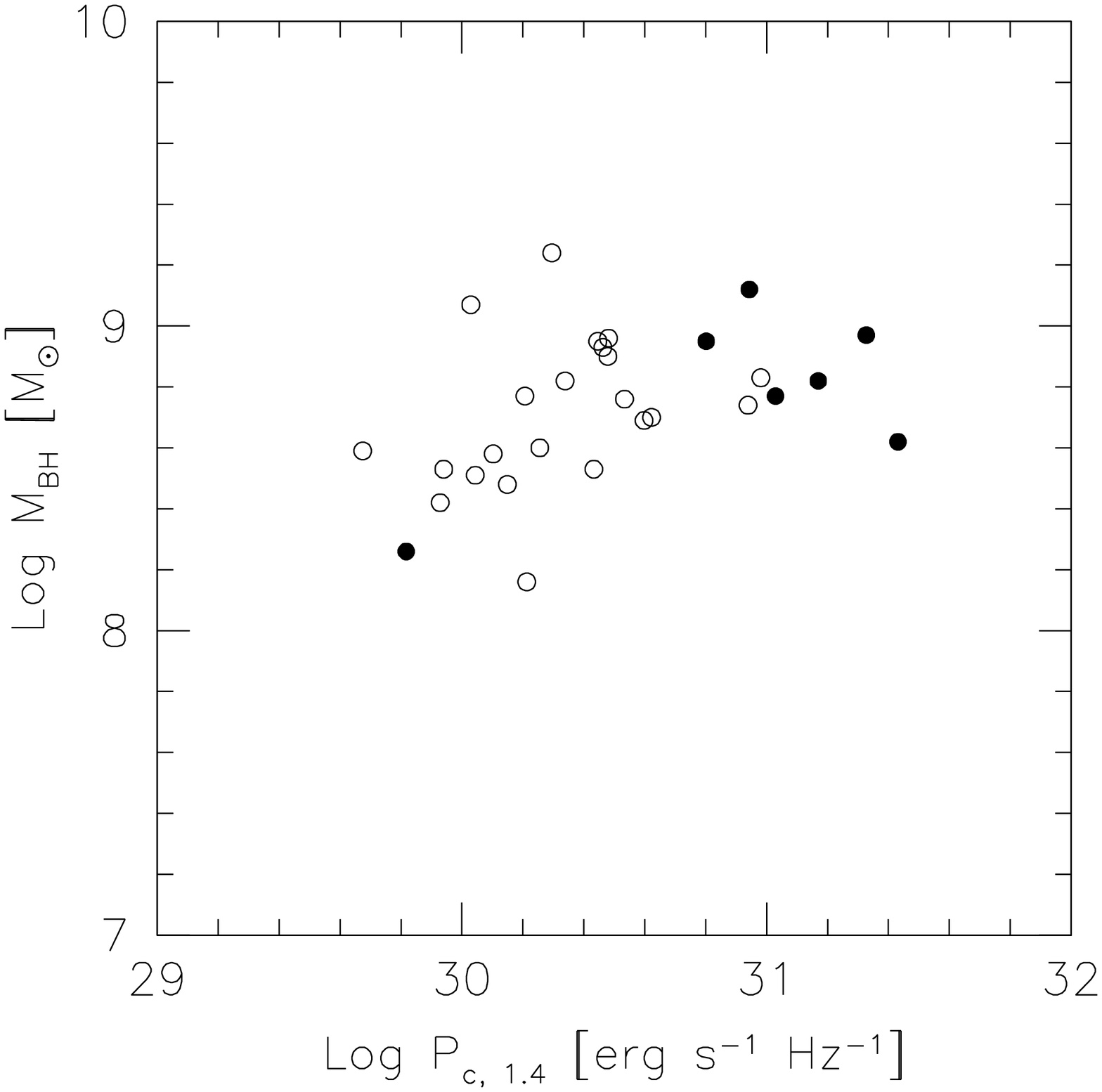}
\figcaption{Black hole mass vs. core radio power at 1.4 GHz. Left: observed
  quantities; right: core luminosity de-beamed assuming
  $\Gamma=1/\sin\theta$. Filled and empty symbols represent HBL and LBL
  respectively. \label{fig:bh2} }
\end{figure}

In Figs.~\ref{fig:bh1} and \ref{fig:bh2} we therefore plot the black hole mass
vs. radio power using the estimates of \citet{fct03}. In Fig.~\ref{fig:bh1}, we
use the observed total radio power, as measured at 325 MHz; in
Fig.~\ref{fig:bh2}, we use the core power at 1.4 GHz: in the left panel we plot
the observational data, in the right one we consider the de-beamed core
luminosity, adopting a Lorentz factor $\Gamma=1/\sin\theta$. Since the
intrinsic core luminosity correlates with the total power, we expect similar
behaviors in Figs.~\ref{fig:bh1} and \ref{fig:bh2} (right).

Neither the total nor the core radio power (corrected and uncorrected)
seem to be good indicators of the black hole mass, as the same BH mass is found
in objects with radio power more than two orders of magnitude apart. Although
weak, a trend of larger $M_{\rm BH}$ for more powerful radio sources is however
present; we also note that the corrected nuclear powers are spread over a
narrow range, so that any possible correlation becomes harder to reveal. We
plan in the near future to add lower power sources to the BL Lacs to obtain a
larger range.

We report in Table \ref{tab:bh} the correlation coefficients for the present
sample, describing the significance of the trends visible in the various
plots. Since both the radio power and the black hole mass (estimated from host
luminosity) include a dependence on redshift, we give in Col. (3) also the
coefficient corrected against $z$ and the corresponding probability of the null
hypothesis in Col. (4). Since this probability is relatively low, it is likely
that a correlation between BH mass and radio power is actually present. The
large scatter in the correlation, however, suggests that black hole mass alone
is not a critical parameter for the emitted radio power in nearby BL Lacs.

\subsection{Radio and optical nuclear luminosity}
\label{sec:core}

\begin{figure}
\plottwo{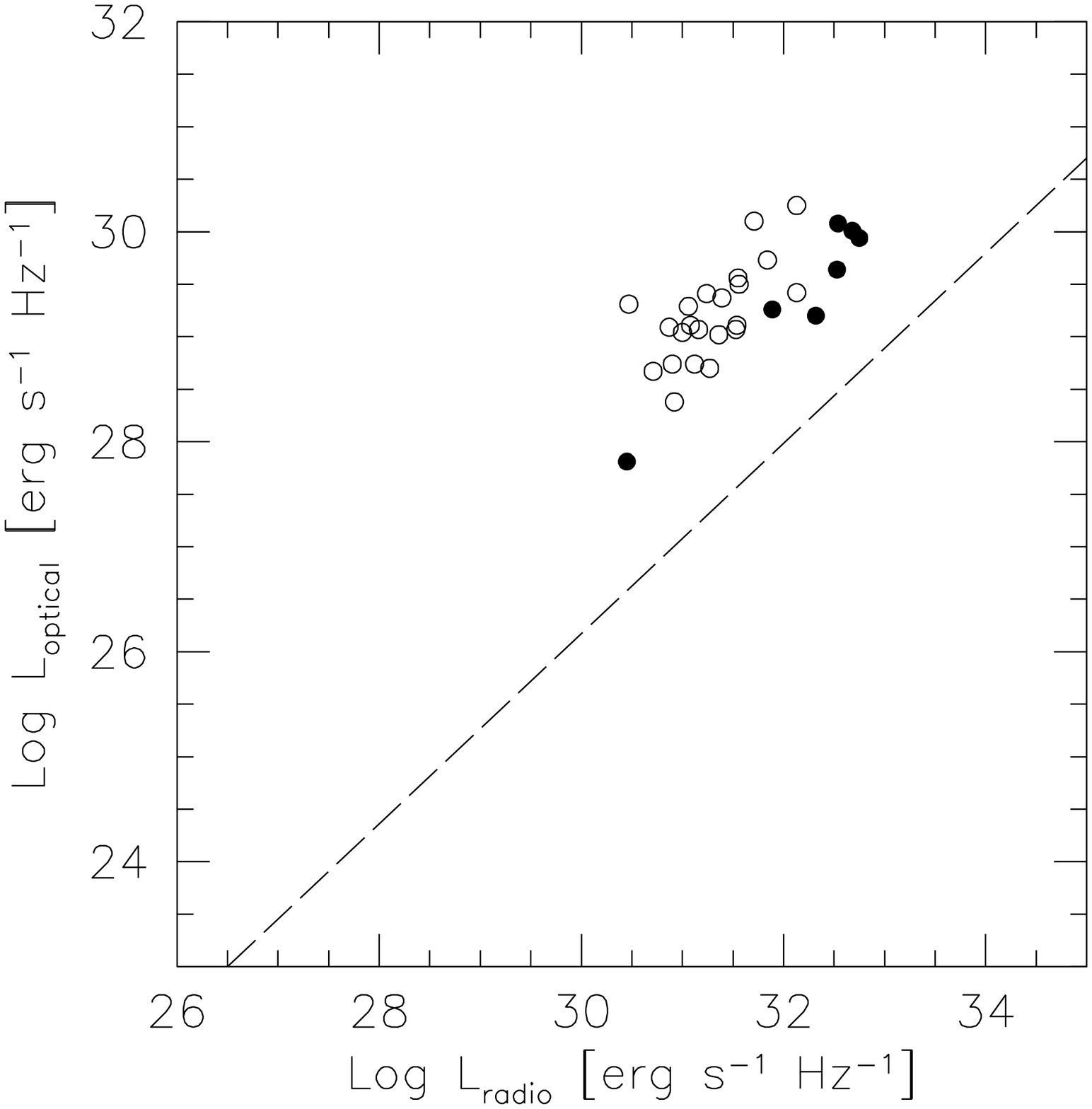}{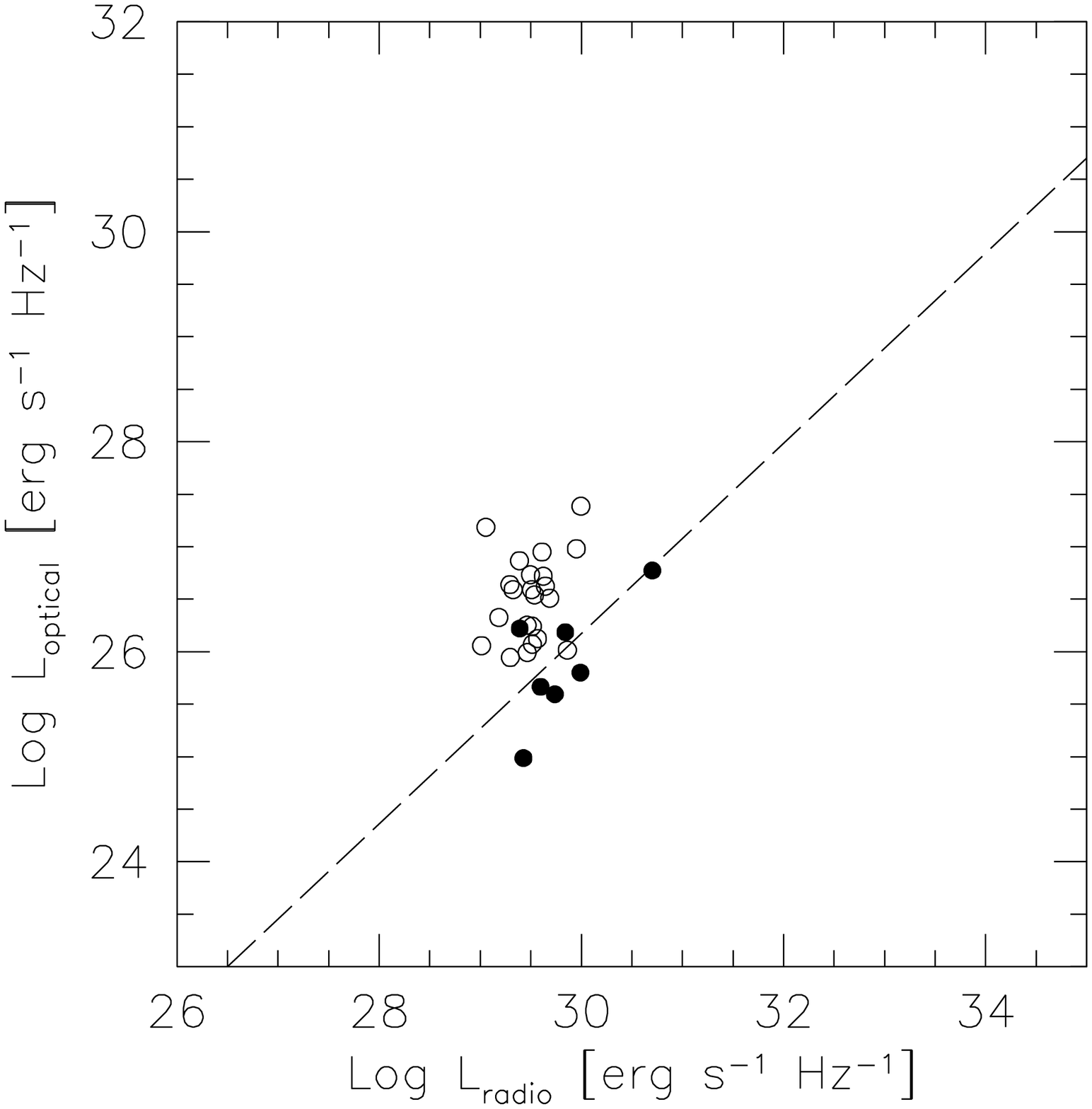}
\figcaption{Optical vs radio core power diagram for BL Lacs in the present
  sample; empty symbols are HBL, filled symbols are LBL. The dashed line shows
  the radio-optical correlation for cores of FR~I radio galaxies
  \citep{chi99}. Left: observed luminosities; right: luminosities corrected for
  Doppler beaming, in the case of a viewing angle of $60^\circ$ and assuming
  $\alpha_O =1$ and $\alpha_R=0.0$ (see \S \ref{sec:core}). \label{fig:optical} }
\end{figure}

\citet{chi99} have exploited Hubble Space Telescope and radio data to discuss
the physical properties of nuclei of low power radio galaxies, finding a linear
correlation between optical and radio luminosity. This correlation is
interpreted in terms of a common non thermal origin. \citet{har00} have
studied the nuclei of 3CRR radio galaxies with similar tools, concluding that
the emission is jet related and that their properties are related to those of
BL Lacs.  Our high quality radio data and the optical HST results
\citep{urr00,fal00} allow us to extend this line of research, moving from the
observational properties found in our sample of low redshift BL Lacs. We
present in Fig.~\ref{fig:optical} the comparison between optical and radio core
luminosity for BL Lacs, together with the correlation found by \citet{chi99}
for FR~I nuclei.

The objects in the present sample span almost three orders of magnitude in
observed radio core luminosity and the optical core luminosity increases
linearly with it over this interval (left panel). A strong correlation
($P>99.99\%$) between optical and radio core luminosity emerges clearly. The
flux densities correlate as well as the powers, confirming that this is not an
artifact introduced by the common dependence on the luminosity distance
\citep[see also the discussion on correlation between intrinsic luminosities
in][]{fer84,fei83}. We show here the luminosities only for the sake of
homogeneity in the comparison with \citet{chi99}. This correlation supports the
suggestion that, as in FR I radio galaxies, the emission in the two wavebands
is ascribed to the same non thermal process, with negligible contribution from,
e.g., a thermal disk. However, the optical cores of BL Lacs are about two
magnitudes brighter than those of FR I galaxies with the same core radio
luminosity. Notice that we are plotting here {\it observed} values,
i.e. quantities that are affected by beaming. While it is interesting to note
that this keeps the correlation tight, we argue that this may be responsible
for the optical offset observed for BL Lacs. \citet{chi00} obtained a similar
result from the BL Lacs in the Slew survey, while the LBL of the 1 Jy sample
result less displaced from the correlation than our objects.

In the right panel, we show the luminosities as corrected for the derived
Doppler factor. We assume for the optical emission a spectral index
$\alpha_O = 1$ \citep{fal94} and for the radio core $\alpha_R = 0.0$ (Table
\ref{tab407}). We also re-calculate the luminosities in the hypothesis of a
viewing angle of $60^\circ$, which is the average angle expected for the
population of FR I radio galaxies used by \citet{chi99} to derive the
correlation. As an effect of de-beaming, all the points move to lower radio and
optical luminosities and approach the correlation for radio galaxies, with some
scatter. This makes a strong point in favor of a common nature of BL Lacs and
FRI radio galaxies cores. The little scatter left in the de-beamed values and
in particular the systematic offset for HBL sources, though little, could be
suggestive of some finer detail. For example, a two velocity jet (fast spine
and slower external layer, with different weights in LBL and HBL) is a
plausible solution that accounts also for other observational properties, such
as limb brightening of jets \citep{gir04a}.

Finally, we note that it would have been possible to use our VLBI data to
calculate parsec scale radio luminosities. We have not done this for the sake
of homogeneity in the comparison with FRIs. However, it will be interesting to
reconsider the subject once we have parsec scale data for a large unbiased
sample of low power radio galaxies, such as the Bologna Complete Sample
\citep[][Giroletti et al. in prep]{gio05}.

\subsection{Ledlow \& Owen diagram}

\begin{figure}
\plottwo{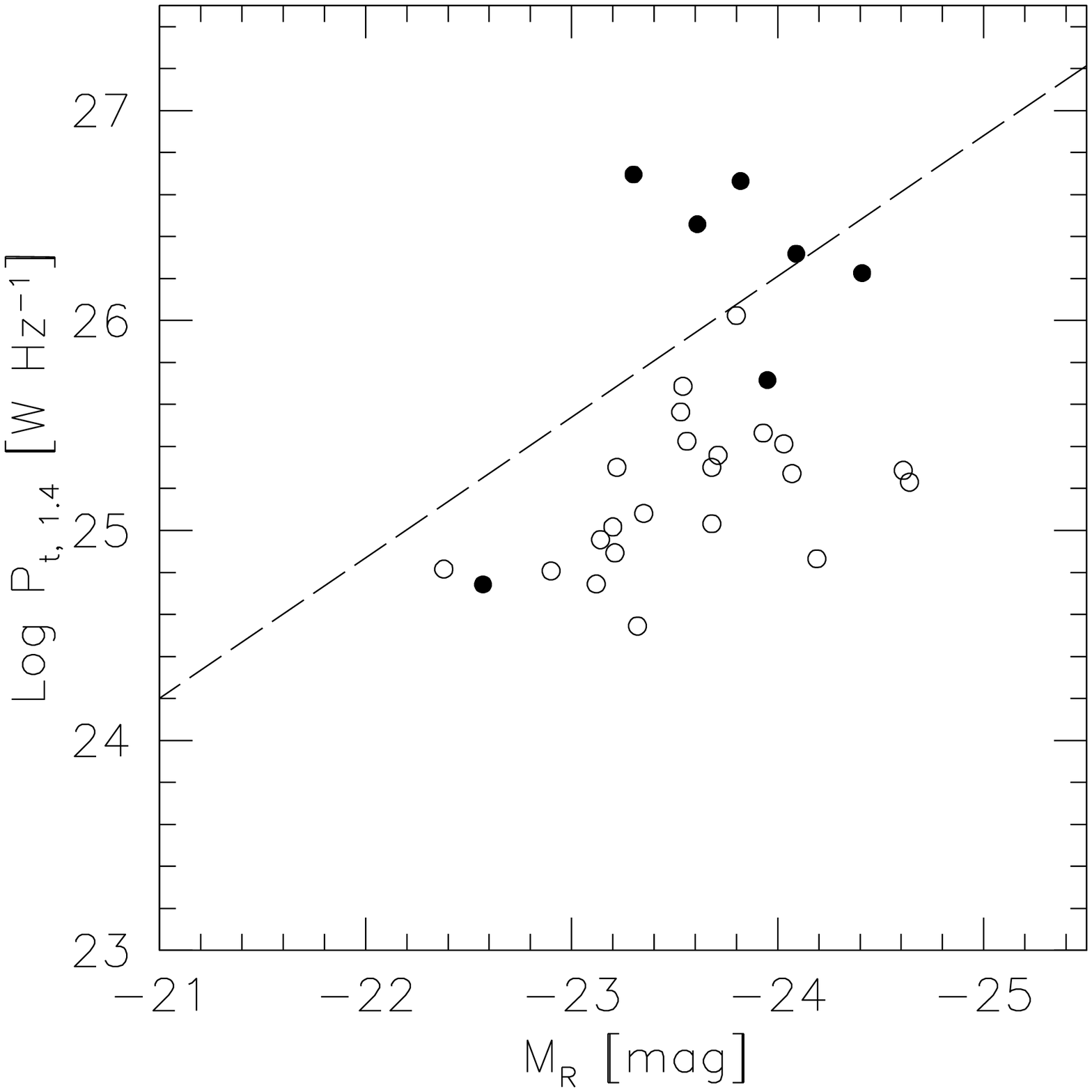}{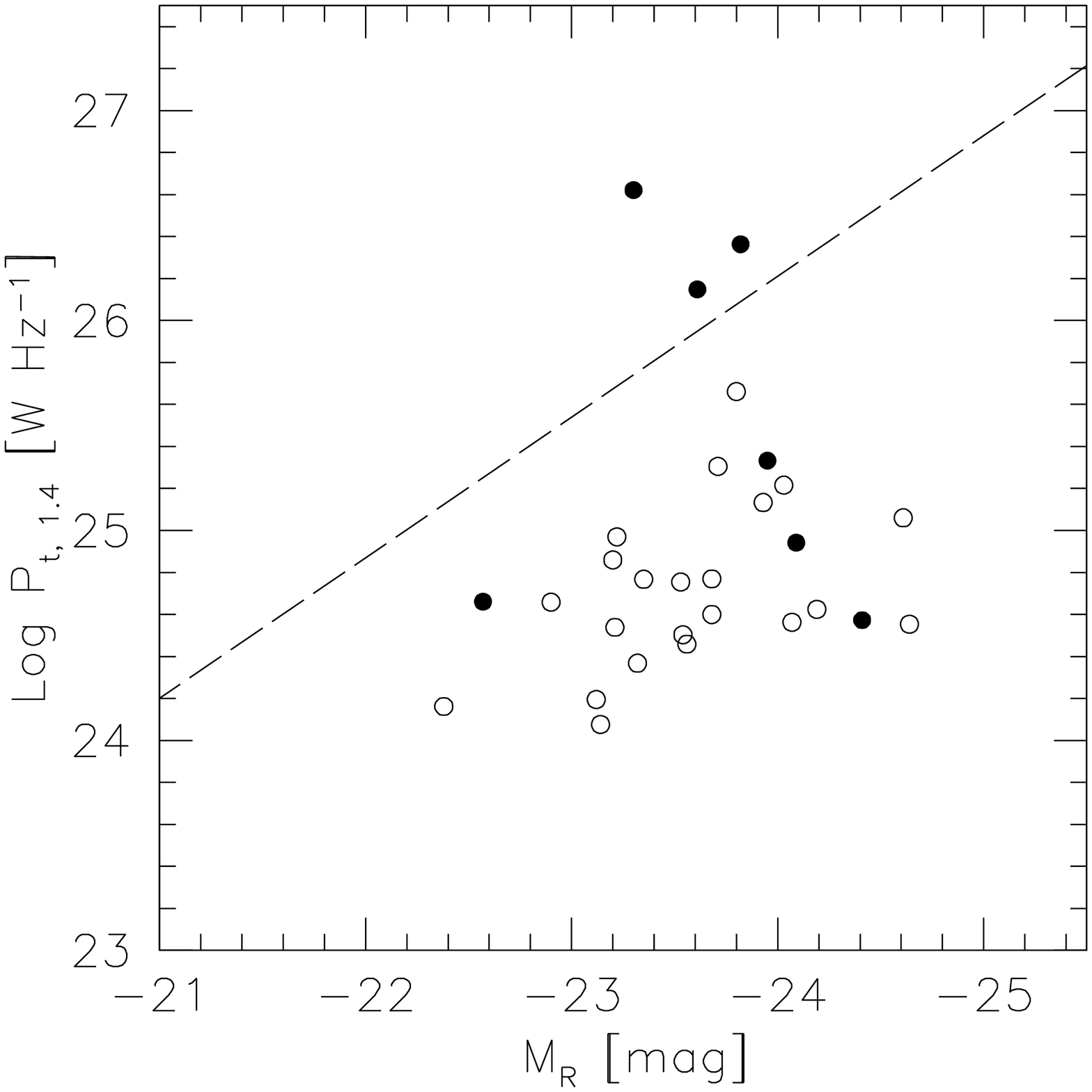}
\figcaption{Ledlow \& Owen diagram; left: observed, right: intrinsic. For ease
  of comparison with similar plots, we have rescaled our data to H$_0 = 50$ km
  s$^{-1}$ Mpc$^{-1}$ and $q_0 = 0$. The quantities on the $x$ and $y$ axis
  (host magnitude and total radio power) are a measure of the black hole mass
  and accretion rate, respectively. The dashed line corresponds to $L_{\rm jet}
  \sim 0.015 L_{\rm Edd}$ \citep{ghi01}. The three objects that remain above
  the line are 0521$-$365, 0829+046, 2200+420. Filled and empty circles
  represent LBL and HBL, respectively. \label{fig:ledlow} }
\end{figure}

Figure \ref{fig:ledlow} shows the distribution of the objects belonging to the
present sample in the Ledlow \& Owen diagram, in which the dashed line
separates FR~I and FR~II radio galaxies \citep{led96}. When we consider the
observed values (left panel), about 20\% of our nearby BL Lacs fall in the FR
II domain or across the transition region. However, the core fraction of the
total luminosity is still significant at 1.4 GHz, so we have to separate the
core and extended radio luminosity and de-beam the former. We show the
intrinsic quantities that we derive in this way in the right panel: most
objects move deep into the region typical of FR I radio galaxies, leaving only
three sources in the FR~II and transition region: 0521$-$365, 0829+046, and
2200+420 (all LBL).

In order to discuss the implication of this distribution in the Ledlow \& Owen
diagram, we consider again the relationship between optical magnitude of the
host and BH mass as well as the one between radio power and total jet kinetic
power \citep{wil99}. We can then interpret the diagram in terms of BH mass and
jet power: for a given $M_\mathrm{BH}$, FR IIs have jets of total power above a
certain value, which becomes larger for more massive black holes. In other
words, there is a critical ratio between jet power and black hole mass that
divides FR I and FR II radio galaxies. Our sources seem to produce radio jets
of low kinetic power with respect to FSRQ and FR II radio galaxies with similar
BH masses.

One can speculate about the interpretation of the FR~I/FR~II dichotomy;
for example, \citet{wil99} have related the total radio power to the narrow
emission line luminosity, which is directly produced by photo-ionization from
the nuclear accreting radiation. Using these correlations, one could estimate
accretion rates for FRIs and BL Lacs and speculate that they are sub-Eddington
systems with sub-critical accretion rates \citep{ghi01}. Unfortunately, line
luminosity is generally not available for sources in our sample and the
validity of such assumptions can not be established.

\subsection{The fundamental plane of BH activity}

A correlation between X-ray and radio emission is expected if there is a
fundamental connection between accretion flows and jet activity. \citet{mer03}
have considered a large sample of objects (AGN and X-ray binaries) powered by
super massive black holes (SMBH) as well as galactic BHs. From the study of
their radio core luminosity, 2-10 keV X-rays luminosity, and BH mass, a
Fundamental Plane (FP) of BH activity is found. However, BL Lacs have been
excluded from this study, since Doppler boosting of the jet emission is
expected to spoil the correlation.

\begin{figure}
\plottwo{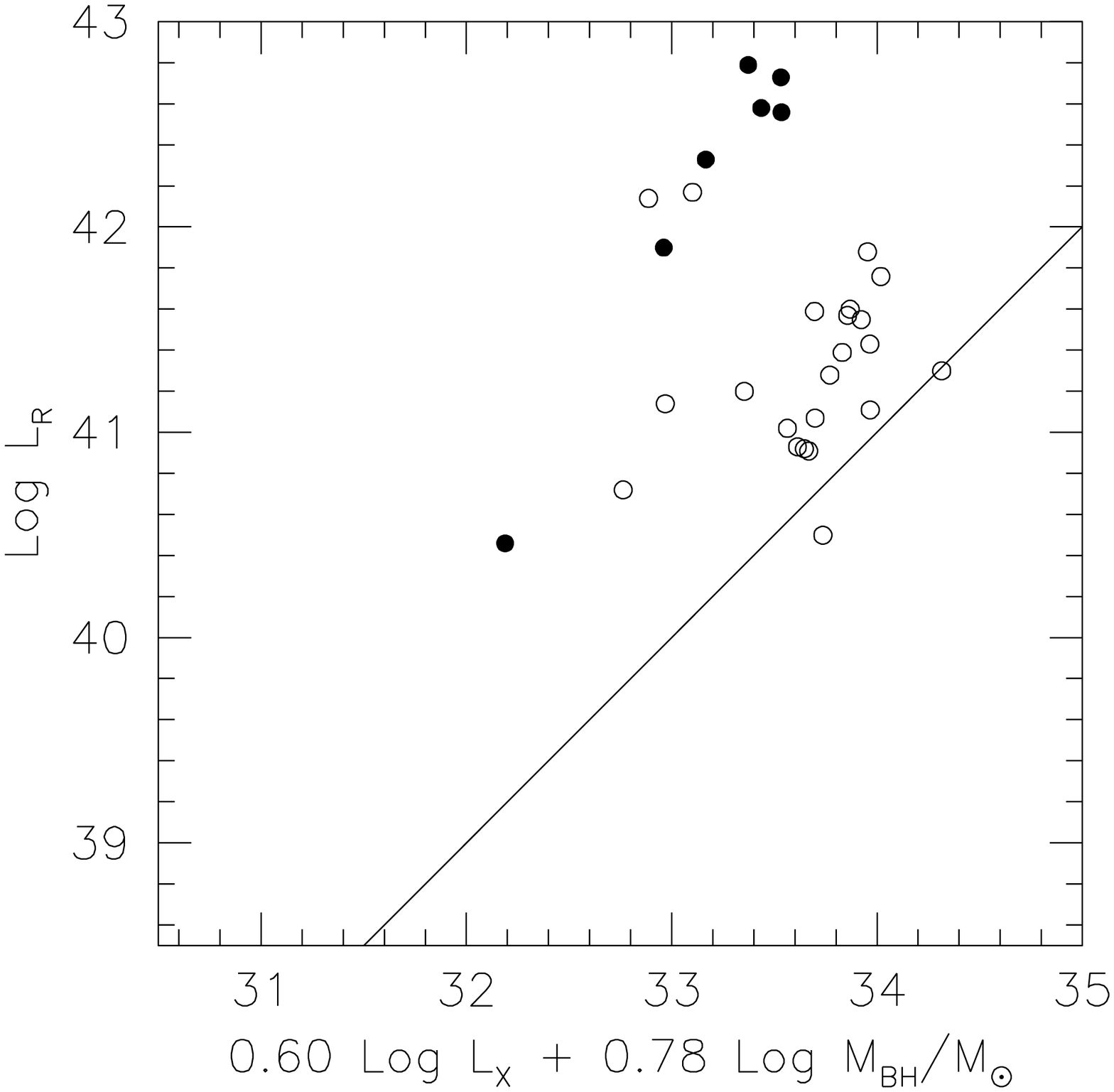}{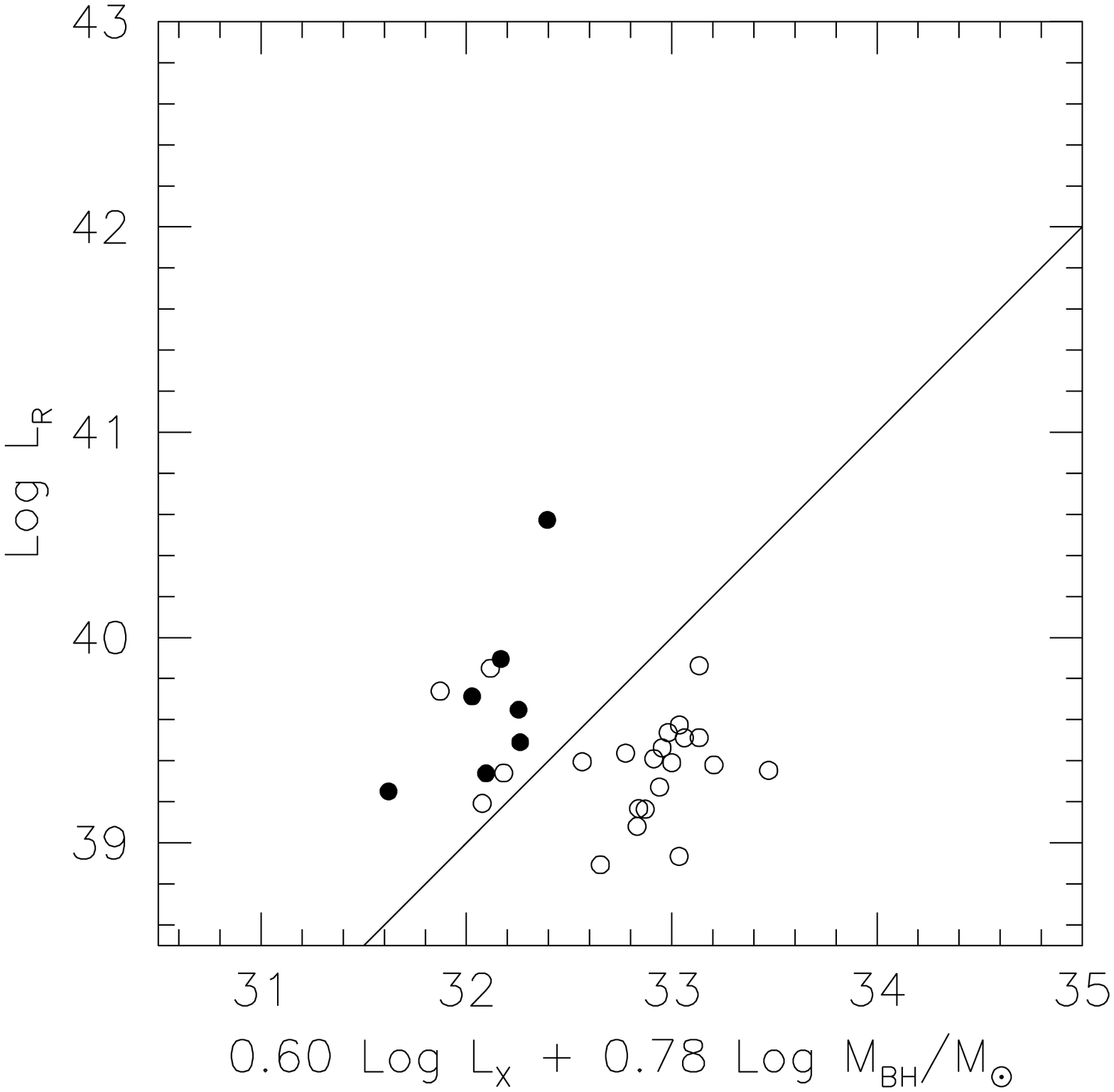}
\figcaption{The ``edge-on'' view of the fundamental plane of BH activity; the
  solid line shows the best-fit to the sample of SMBH and galactic BHs studied
  by \citet{mer03}. Left: observed quantities; right: corrected for Doppler
  boosting and re-calculated for a mean viewing angle of $60^\circ$. Filled and
  empty circles represent LBL and HBL, respectively. \label{fig:fp} }
\end{figure}

In Fig.~\ref{fig:fp} (left panel), we show the position of our nearby BL Lacs
in the projection of the FP, as obtained from the observed (i.e.~Doppler
boosted) values: it is readily visible that the observed radio luminosities are
well above the correlation. From the derived intrinsic de-beamed
values\footnote{The X-rays luminosities have been de-beamed assuming a spectral
index $\alpha_X = 0.7$}, we then plot in the right hand panel the new $L_R$
vs. $0.60 \mathrm{Log} L_X + 0.78 \mathrm{Log} M_\mathrm{BH}/M_{\odot}$,
recalculated in the case of a viewing angle of 60$^\circ$ (which is the average
angle for the objects used to derive the correlation).  The corrected data seem
to cluster around the FP relation, albeit with a significant scatter in the
points. Although no systematic general offset is found, we note that LBLs
(filled circles) lie above/left of the correlation, while HBLs are
preferentially found below/right of it.

It is thus interesting that AGN with radio jets, as our BL Lacs, do follow the
same relation of the bulk of SMBHs and galactic BHs. On one hand, it could
support the existence of the general fundamental connection between accretion
flows and jet activity \citep{mer03,fal04}. In this case, the possible offset
between LBLs and HBLs could be suggestive of possible refining of the proposed
models, requiring more powerful radio jets in LBLs or accreting process that is
more radiatively efficient in HBLs. On the other hand, it has been suggested
that the FP relation is actually an artifact due to the flux-limited nature of
the sample considered \citep{bre05}. Once the Doppler-boosting dependence has
been removed, a common dependence on distance in the plotted quantities could
become apparent.  Since our sample is volume-limited ($z<0.2$), rather than
flux-limited, we should however be free from this bias: we have radio and X-ray
fluxes spanning 3 orders of magnitude each (radio fluxes between 7 mJy and 5
Jy, and X-rays 2-10 keV fluxes between 0.4 and 240 erg cm$^{-2}$ s$^{-1}$).

\section{CONCLUSIONS}
\label{conclusions}

We have presented new high resolution images for 9 BL Lacs. Parsec-scale images
reveal an almost ubiquitous presence of one-sided jets. Under the assumption of
intrinsically symmetric jets, this provides evidence for relativistic
velocities in the parsec scale region of jets. This result supports the
previous findings on the same sample \citep{gir04b}, and we conclude that the
Lorentz factors in the sample cluster around $\Gamma \sim 4$, mainly on the
basis of the core dominance argument. In the best studied sources, thanks to
the short baselines provided by the MERLIN, we study the jet out to several
hundred parsecs, and still find no evidence of counter-jets. On kiloparsec
scales, the structure of 1728+502 remains aligned with the inner region
direction, while 1215+303 and Mrk~421 reveal a symmetric structure without a
clear connection with the parsec scale jet. In Mrk~421 we have evidence of limb
brightening of the jet, from the very inner jet region ($\sim 2$ mas from the
core) out to $\sim 30$ mas. This result, similar to what is found in the other
TeV source Mrk~501 \citep{gir04a}, is of great importance in the development of
jet launch models and the study of the jet/environment interaction.

X-ray fluxes and high resolution optical data available for our whole sample of
29 nearby BL Lacs allow us to perform a multi-wavelength analysis. In
particular, by resolving the host galaxy and separating the non-thermal core,
we have been able to discuss the physical properties of the BL Lac
phenomenon. Given the distance-limited nature of the present sample, we are
confident of the robustness of our results, which we summarize as follows:

1. There is a significant correlation between the mass of the central SMBH and
   the radio luminosity, both nuclear and extended, independent of any common
   distance dependence. However, there is a large scatter in the correlation,
   so that radio luminosity alone can not be taken as an indicator of the BH
   mass.

2. Radio and optical nuclear luminosities (and fluxes) show a strong
   correlation, indicating a common origin for the emission at both
   wavelengths, i.e.~non thermal synchrotron radiation. Moreover, if we de-beam
   the observed luminosities, the cores follow with little scatter the same
   correlation found for low power radio galaxies by \citet{chi99}, in strong
   support of the unification of BL Lacs and FR~Is based on orientation.

3. Still in support of the BL Lac/FR~I unification, our sources lie below the
   dividing line in the Ledlow \& Owen diagram; for any given
   $M_\mathrm{BH}$, low redshift BL Lacs have lower radio power than FR II
   radio galaxies of similar $M_\mathrm{BH}$.

4. The radio and X-ray luminosities do not follow the fundamental plane of BH
   activity found for other super-massive and galactic BH, clearly because of
   the relevant Doppler boosting of the jet. When we consider de-beamed
   quantities, the FP relation becomes quite well-reproduced, although some
   overall scatter and LBL-HBL offset appear.

\begin{acknowledgments} 
 
We thank an anonymous referee for a constructive report that improved the
clarity of the paper. This research has made use of the NASA/IPAC Extragalactic
Database (NED) which is operated by the Jet Propulsion Laboratory, Caltech,
under contract with NASA and of NASA's Astrophysics Data System (ADS)
Bibliographic Services. The European VLBI Network is a joint facility of
European, Chinese, South African and other radio astronomy institutes funded by
their national research councils. MERLIN is a National Facility operated by the
University of Manchester at Jodrell Bank Observatory on behalf of PPARC.  The
National Radio Astronomy Observatory is a facility of the National Science
Foundation operated under cooperative agreement by Associated Universities,
Inc. This material is based upon work supported by the Italian Ministry for
University and Research (MIUR) under grant COFIN 2003-02-7534.

\end{acknowledgments}

\clearpage

\begin{deluxetable}{llrrrccrrcr}
\tabletypesize{\scriptsize}
\tablecaption{Objects in the Sample \label{table1}}
\tablehead{ 
\colhead{Name} & \colhead{Name} & \colhead{$z$} & \colhead{$\mathrm{Log} P_\mathrm{1.4\,GHz}$} & \colhead{$-M_R$} & \colhead{$\mathrm{Log} L_\mathrm{2-10 keV}$} & \colhead{Ref.} & \colhead{$\theta$} & \colhead{$\delta$} & \colhead{Class} & \colhead{Sample} \\
\colhead{(IAU)} & \colhead{(Other)} & \colhead{} & \colhead{(W Hz$^{-1}$)} & \colhead{(mag)} & \colhead{(erg s$^{-1}$)} & \colhead{} & \colhead{($^\circ$)} & \colhead{} & \colhead{} & \colhead{} \\
\colhead{(1)} & \colhead{(2)} & \colhead{(3)} & \colhead{(4)} & \colhead{(5)} & \colhead{(6)} & \colhead{(7)} & \colhead{(8)} & \colhead{(9)} & \colhead{(10)} & \colhead{(11)} }
\startdata
0229+200  &       & 0.140 & 24.62 & 23.96 & 45.2 & 1 & 19 & 3.1 & H & HEAO-A3 \\
0347$-$121&       & 0.188 & 24.32 & 22.56 & 45.0 & 2 & 23 & 2.6 & H & HEAO-A3 \\
0350$-$371&       & 0.165 & 24.40 & 22.69 & 44.4 & 3 & 21 & 2.8 & H & EMSS \\
0521$-$365&       & 0.055 & 26.07 & 22.60 & 44.1 & 2 & 21 & 2.6 & L & HEAO-A3 \\
0548$-$322&       & 0.069 & 24.72 & 23.05 & 44.6 & 2 & 32 & 1.9 & H & HEAO-A2 \\
0706+591  &       & 0.125 & 24.75 & 23.38 & 44.7 & 4 & 20 & 2.9 & H & HEAO-A3 \\
0806+524  &       & 0.137 & 24.90 & 22.87 & 45.3 & 1 & 14 & 4.1 & H & Slew \\
0829+046  &       & 0.180 & 25.97 & 23.18 & 44.4 & 4 & 10 & 5.7 & L & HEAO-A3 \\
0927+500  &       & 0.188 & 24.26 & 22.50 & 45.3 & 1 & 16 & 3.5 & H & Slew \\
1101+384  &Mrk~421& 0.031 & 24.27 & 22.57 & 45.0 & 2 & 19 & 3.0 & H & HEAO-A3 \\
1133+704  &Mrk~180& 0.046 & 24.17 & 22.24 & 43.7 & 2 & 24 & 2.5 & H & HEAO-A3 \\
1212+078  &       & 0.136 & 24.80 & 23.26 & 44.6 & 1 & 17 & 3.4 & H & Slew \\
1215+303  &       & 0.130 & 25.36 & 23.15 & 43.7 & 2 & 15 & 3.7 & H & Slew \\
1218+304  &       & 0.182 & 24.74 & 22.92 & 45.4 & 2 & 16 & 3.6 & H & HEAO-A2 \\
1229+643  &       & 0.164 & 24.59 & 23.43 & 44.8 & 3 & 17 & 3.3 & H & EMSS \\
1255+244  &       & 0.141 & 23.88 & 22.67 & 45.1 & 2 & 24 & 2.5 & H & Slew \\
1418+546  &OQ~530 & 0.152 & 25.64 & 23.44 & 44.0 & 2 & 11 & 5.1 & L & PG \\
1426+428  &       & 0.129 & 24.36 & 23.03 & 45.2 & 2 & 21 & 2.7 & H & HEAO-A3 \\
1440+122  &       & 0.162 & 24.62 & 23.04 & 45.1 & 1 & 18 & 3.3 & H & Slew \\
1514$-$241&AP~Lib & 0.049 & 25.06 & 22.88 & 43.4 & 4 &  6 & 9.3 & H & 1 Jy \\
1652+398  &Mrk~501& 0.034 & 24.61 & 24.01 & 44.9 & 2 & 17 & 3.5 & H & 1 Jy \\
1728+502&I\,Zw 187& 0.055 & 24.19 & 21.68 & 44.3 & 1 & 21 & 2.8 & H & HEAO-A3 \\
1807+698  &3C~371 & 0.051 & 25.08 & 23.25 & 43.3 & 2 & 17 & 3.5 & L & 1 Jy \\
1959+650  &       & 0.048 & 24.12 & 22.46 & 44.8 & 2 & 17 & 3.4 & H & HEAO-A3 \\
2200+420  &BL~Lac & 0.070 & 25.82 & 22.95 & 44.5 & 2 &  9 & 6.7 & L & 1 Jy \\
2201+044  &       & 0.027 & 24.12 & 21.86 & 42.9 & 4 & 29 & 2.1 & L & HEAO-A3 \\
2254+074  &       & 0.190 & 25.53 & 23.77 & 43.9 & 4 &  9 & 6.2 & L & 1 Jy \\
2344+514  &       & 0.044 & 24.24 & 23.53 & 44.2 & 2 & 21 & 2.8 & H & Slew \\
2356$-$309&       & 0.165 & 24.62 & 22.56 & 45.5 & 2 & 19 & 3.0 & H & HEAO-A3 \\
\enddata
\tablecomments{Col. 7: reference for the X-ray luminosity; (1)
  \citet{per96}, (2) \citet{don05}, (3) \citet{rec00}, (4)
  \citet{don01}. Col. 8: uncertainty on the viewing angle $\theta$ is
  typically $\pm 5^\circ$ (but see discussion in
  \S~\ref{discussion}). Col. 10: H -- High frequency peaked BL Lac, L -- Low
  frequency peaked BL Lac. Col. 11: 1 Jy \citep{sti91}; HEAO-A2 \citep{pic82};
  HEAO-A3 \citep{rem99}; EMSS \citep{mor91}; Slew \citep{sch93,per96}}
\end{deluxetable}

\begin{deluxetable}{llll}
\tablecaption{Log of observations considered in this paper\label{tab2}}
\tablehead{ 
\colhead{} & \colhead{} & \colhead{Freq.} & \colhead{} \\
\colhead{Date} & \colhead{Array} & \colhead{(GHz)} & \colhead{sources}
}
\startdata
2002/6/7 & E   & 1.6 & 0706+591, 0806+524, 1133+704, 1215+303, \\
         &     &     & 1218+304, 1229+643, 1426+428, 1728+502 \\
2002/5/26 & E, M, EM & 5 & 1215+303, 1728+502 \\
1997/2/17 & E, M, EM & 5 & 1101+384 \\
1996/2/10 & G & 1.6 & 1101+384 \\
1995/7/22 & V & 5 & 1101+384 \\
\enddata		     
\tablecomments{Col. (2): E = EVN; M = MERLIN; EM = EVN+MERLIN; G = Global VLBI; V = VLBA + Y1 (see text for details).}			     
\end{deluxetable}

\begin{deluxetable}{llrr}
\tablecaption{EVN Observations at 1.6 GHz, Image Parameters\label{imagedata}}
\tablehead{
\colhead{Name} & \colhead{HPBW} & \colhead{Noise ($3\sigma$)} & \colhead{Peak} \\
\colhead{} & \colhead{(mas $\times$ mas, $^\circ$)} & 
\multicolumn{2}{c}{(mJy beam$^{-1}$)}
}
\startdata
 0706+591 & $8.8 \times 6.9, \,-55$   & 0.5 & 33.5 \\
 0806+524 & $10.9 \times 4.3,\, -26$ & 2.0 & 98.0       \\
 1133+704 & $5.1 \times 4.5, \,-28$ & 3.0 & 67.2 \\
 1215+303 & $12.5 \times 4.1,\, 7$ & 1.0 & 224.5 \\
 1218+304 & $12.5 \times 4.2,\, 6$ & 1.0 & 48.7 \\
 1229+643 & $10.0 \times 4.6,\, 7$ & 1.3 & 44.1 \\
 1426+428 & $8.8 \times 4.3, \,3$ & 1.4 & 28.8 \\
 1728+502 & $12.1 \times 5.5,\, -58$ & 1.0 & 107.4 \\
\enddata
\end{deluxetable}

\begin{deluxetable}{lllrr}
\tablecaption{EVN+MERLIN Observations at 5 GHz, Image Parameters 
  \label{merlindata}}
\tablehead{
\colhead{Source} & \colhead{Array} & \colhead{HPBW} & \colhead{Noise ($3\sigma$)} & \colhead{Peak} \\ 
\colhead{} & \colhead{} & \colhead{(mas $\times$ mas, $^\circ$)} & \multicolumn{2}{c}{(mJy beam$^{-1}$)}
}
\startdata
1215+303 & MERLIN     & $55   \times 52,  \,   29$  & 0.35 & 315 \\
         & EVN+MERLIN & $11.7 \times 8.2, \, 12$  & 0.40 & 294 \\
         & EVN        & $5.2  \times 1.9, \,  7$   & 0.38 & 274 \\
1728+502 & MERLIN     & $51   \times 47,  \,   -31$ & 0.40 & 163 \\
         & EVN+MERLIN & $7.6  \times 6.4, \,  38$  & 0.27 & 116 \\
         & EVN        & $2.2  \times 1.6, \,  15$  & 0.43 &  92 \\
Mrk 421  & MERLIN     & $79   \times 47,  \,   11$ & 1.2 & 530 \\
         & EVN+MERLIN & $11.4 \times 9.8, \, 43$ & 0.87 & 426 \\
         & EVN        & $6.5  \times 3.1, \,  -9$ & 0.65 & 407 \\
         & VLBA+Y1    & $3.6  \times 2.6, \,  -19$ & 0.64 & 338 \\
         & Global VLBI &$7.0  \times 3.9, \, 8$ & 1.2 & 332 \\
\enddata
\end{deluxetable}

\begin{deluxetable}{llrrrr}
\tablecaption{Results of the Model Fits to the EVN 1.6 GHz Observations \label{modelfits}}
\tablehead{
\colhead{} & \colhead{} & \colhead{Flux} \\
\colhead{Source} & \colhead{Comp.} & \colhead{density} & \colhead{$r$} & \colhead{$\theta$} & \colhead{$b$} \\
\colhead{} & \colhead{} & \colhead{(mJy)} & \colhead{(mas)} & \colhead{($^\circ$)} & \colhead{(mas)} \\
\colhead{(1)} & \colhead{(2)} & \colhead{(3)} & \colhead{(4)} & \colhead{(5)} & \colhead{(6)} 
} 
\startdata
0706+591 & core & $ 36\pm4 $ &  ref. &  ref. & $ \le 2.5 $\\
	 & J1   & $  7\pm1 $ &  16.8 &$-$151 & $ 11\pm2 $\\
\hline		  	     	     	       	       	   
0806+524 & core & $ 97\pm10$ &  ref. &  ref. & $ \le 1.8$ \\
	 & J1   & $ 84\pm9 $ &   4.8 &    42 & $ 3.1\pm0.5 $\\
\hline		  	     	     	       	       	   
1133+704 & core & $ 82\pm8 $ &  ref. &  ref. & $ \le 1.8 $\\
	 & J2   & $ 36\pm5 $ &   6.1 &    69 & $ 9.3\pm1.5 $\\
	 & J1   & $  9\pm3 $ &  33.5 &    76 & $ 14 \pm 3 $\\
\hline		  	      	     	       	       	   
1215+303 & core & $261\pm26$ &  ref. &  ref. & $ \le 2.2$ \\
	 & J2   & $ 33\pm3 $ &  18.4 &   143 & $ 9.8\pm1.0 $\\
	 & J1   & $ 25\pm3 $ &  52.7 &   145 & $ 16\pm2 $\\
\hline		  	      	     	       	       	   
1218+304 & core & $ 49\pm5 $ &  ref. &  ref. & $ \le 1.7 $\\
	 & J1   & $ 15\pm2 $ &   5.2 &    84 & $ 4.3\pm1.3 $\\
\hline		  	      	     	       	       	   
1229+643 & core & $ 48\pm5 $ &  ref. &  ref. & $ \le 2.1 $\\
	 & J1   & $ 15\pm2 $ &  23.9 & $-$50 & $ 32\pm10 $\\
\hline		  	      	     	       	       	   
1426+428 & core & $ 32\pm4 $ &  ref. &  ref. & $ \le 1.9 $\\
\hline		  	      	     	       	       	   
1728+502 & core & $110\pm11$ &  ref. &  ref. & $ \le 2.1$ \\
	 & J2   & $ 76\pm8 $ &   7.1 & $-$49 & $ 3.6\pm0.5 $\\
	 & J1   & $ 11\pm2 $ &  21.3 & $-$55 & $ 6.6\pm1.5 $\\
\enddata
\tablecomments{Results of circular Gaussian model fits to the EVN 1.6 GHz visibility data. Col. 2: jet components are numbered J1, J2 from the outermost inward. Columns 4, 5, 6: $r, \theta, b = $~radius, P.A.~and FWHM of the component. See \S\ref{results} for a discussion of uncertainties.}
\end{deluxetable}

\begin{deluxetable}{lrrrrr}
\tablecaption{Core and jet observational properties \label{tab407}}
\tablehead{
\colhead{Source} & \colhead{$S_\mathrm{c, 1.6 GHz}$} & \colhead{$S_\mathrm{c, 5 GHz}$} & \colhead{spectral} & \colhead{$B_\mathrm{j}$} & \colhead{$R_\mathrm{min}$} \\
\colhead{} & \colhead{(mJy)} & \colhead{(mJy)} & \colhead{index} & \colhead{(mJy beam$^{-1}$)} & \colhead{} \\
\colhead{(1)} & \colhead{(2)} & \colhead{(3)} & \colhead{(4)} & \colhead{(5)} & \colhead{(6)} 
}
\startdata
0706+591 & 36  & 36   &  0.0 & 2.0 &  12  \\
0806+524 & 97  & 100  &$-0.0$& 69.5 & 108 \\
1101+384 & 325 & 323  &  0.0 & 22.0 & 110 \\
1133+704 & 82  &  83  &$-0.0$& 32.0 & 33  \\
1215+303 & 261 & 292  &$-0.1$& 9.4 & 150  \\
1218+304 & 49  & 43   &  0.1 & 6.0 & 18   \\
1229+643 & 48  & 26   &  0.5 & 2.0 & 5 \\
1426+428 & 32  &  19  &  0.5 & \nodata & \nodata \\
1728+502 & 110 & 117  &$-0.1$& 18.0 & 200 \\
\enddata
\tablecomments{Core flux densities and spectral indexes are calculated
  considering only $(u,v)$ data in the range $0-40$ M$\lambda$. The 5 GHz data
  are taken from \citetalias{gir04b}, except for 1215+303 and 1728+502 (EVN,
  present work) and 1133+704 and 1426+428 \citep[EVN,][]{kol96}. Flux densities
  at 1.6 GHz are from EVN 2002 data, except for 1101+384 which is from the
  global VLBI experiment of 1996. }
\end{deluxetable}

\begin{deluxetable}{llrrrr}
\tablecaption{Results of the Model Fits to Markarian 421 \label{tab421}}
\tablehead{
\colhead{} & \colhead{} & \colhead{Flux} \\
\colhead{Epoch} & \colhead{Comp.} & \colhead{density} & \colhead{$r$} & \colhead{$\theta$} & \colhead{$b$} \\
\colhead{} & \colhead{} & \colhead{(mJy)} & \colhead{(mas)} & \colhead{($^\circ$)} & \colhead{(mas)} \\
\colhead{(1)} & \colhead{(2)} & \colhead{(3)} & \colhead{(4)} & \colhead{(5)} & \colhead{(6)} 
} 
\startdata
1995.56 & Core & $335\pm34$ & ref. & ref. & $\le0.4$ \\
     & C5   & $25\pm3$  &   1.8 & $-$32  & $\le0.4$ \\
     & C4   & $15\pm2$  &   4.7 & $-$38  & $1.6\pm0.5$ \\
     & C3   & $81\pm8$  &  27.6 & $-$56  & $38\pm15$ \\
1996.11 & Core & $232\pm23$ & ref. & ref. & $\le 0.4$ \\
     & C5   & $105\pm11$ &   1.3 & $-$40  & $\le 0.4$ \\
     & C4   & $43\pm5$   &   5.1 & $-$39  & $3.0\pm0.7$ \\
     & C3   & $219\pm22$ &  26.2 & $-$55  & $48\pm15$ \\
1997.13 & Core & $398\pm40$ & ref. & ref. & $\le 0.4$ \\
     & C5   & $24\pm3$  &   1.4 & $-$42  & $\le 0.4$ \\
     & C4   & $19\pm2$  &   5.7 & $-$36  & $ 1.3\pm0.5$ \\
     & C3   & $17\pm2$  &  16.4 & $-$39  & $ 10\pm5$ \\
\enddata
\tablecomments{Results of circular Gaussian model fits to the visibility data of Mrk~421. C5 and C4 are labeled following \citet{pin05}; the more extended C3 is resolved in their high frequency observations and is also poorly constrained in our models. Columns 4, 5, 6: $r, \theta, b = $~radius, P.A.~and FWHM of the component. Uncertainty on $r$ is typically 0.2 mas for C5, 0.5 mas for C4.}
\end{deluxetable}

\begin{deluxetable}{lrrr}
\tablecaption{Correlation coefficient for BH mass - radio power \label{tab:bh}}
\tablehead{
\colhead{Radio power} & \colhead{$r_z$} & \colhead{$r_{\mathrm{no}-z}$} & \colhead{$P_\mathrm{null}$}   \\
\colhead{(1)} & \colhead{(2)} & \colhead{(3)} & \colhead{(4)}
}
\startdata
extended                                 & 0.43 & 0.42 & 0.011 \\
core, observed                           & 0.45 & 0.41 & 0.014 \\
core, de-beamed with $\Gamma = 1/\theta$ & 0.42 & 0.38 & 0.021 \\
\enddata
\end{deluxetable}

\clearpage

\end{document}